\documentclass{article}
\usepackage[utf8]{inputenc}

\usepackage{algorithm2e}
\usepackage{xcolor}

\SetAlFnt{\footnotesize}

\SetAlCapSty{xAlCapSty}

\SetCommentSty{xCommentSty}

\SetNlSty{mynlfont}{}{} 

\LinesNumbered

\SetSideCommentRight

\DontPrintSemicolon

\RestyleAlgo{algoruled}

\usepackage[driverfallback=dvipdfm,
bookmarks=true,    
bookmarksopen=true,
bookmarksopenlevel=\maxdimen,
bookmarksdepth=10
]{hyperref}
\hypersetup{
pdfcenterwindow=true, 
pdfpagelabels=true,
pagebackref=true,            
hypertexnames=true,
plainpages=false,
unicode=false,     
pdftoolbar=true,                
    pdfmenubar=true,          
    pdffitwindow=false,         
    pdfstartview={FitH},        
    pdftitle={On Third-Order Accuracy of MUSCL Scheme},
    pdfauthor={Hiroaki Nishikawa},    
    pdfsubject={On Third-Order Accuracy of MUSCL Scheme},       
    pdfcreator={Hiroaki Nishikawa},   
    pdfproducer={Hiroaki Nishikawa}, 
    baseurl={http://www.hiroakinishikawa.com},
    pdfkeywords={}, 
    pdfnewwindow=true,     
    colorlinks=true,       
    linkcolor=black,       
    citecolor=black,       
    filecolor=black,       
    urlcolor=black,        
  breaklinks=true,
  hyperfigures=true,
  backref=true,
breaklinks=false, 
pdfpagelabels,      
pagebackref,        
hypertexnames=true, 
plainpages=false,   
naturalnames        
}
\usepackage[all]{hypcap}

\usepackage[thinlines]{easytable}

\usepackage{color}      
\usepackage{placeins}
\usepackage{indentfirst}
\usepackage{multirow}

\usepackage{amssymb}
\usepackage{amsmath,latexsym}
\usepackage{url}
\usepackage{enumerate}
\usepackage{graphicx}

\usepackage{booktabs}

\usepackage{subcaption}
\usepackage{cleveref}
\usepackage{mathtools}

\usepackage{blindtext,graphicx}
\usepackage[absolute]{textpos}
\begin{document}



\title{\bf A Truncation Error Analysis of Third-Order MUSCL Scheme for Nonlinear Conservation Laws}

 \author{
{Hiroaki Nishikawa}\thanks{Associate Research Fellow ({hiro@nianet.org})}\\
  {\normalsize\itshape National Institute of Aerospace,
 Hampton, VA 23666, USA} 
}

\date{\empty}
\maketitle

\begin{abstract} 
This paper is a rebuttal to the claim found in the literature that the MUSCL scheme cannot be third-order accurate for nonlinear
conservation laws. We provide a rigorous proof for third-order accuracy of the MUSCL scheme 
based on 
 a careful and detailed truncation error analysis. Throughout the analysis, the distinction between the cell average and the point value 
will be strictly made for the numerical solution as well as for the target operator.
 It is shown that the average of the solutions reconstructed at a face by Van Leer's $\kappa$-scheme recovers a cubic solution exactly with $\kappa=1/3$, 
the same is true for the average of the nonlinear fluxes evaluated by the reconstructed solutions, and a dissipation term is already sufficiently small with 
a third-order truncation error. 
Finally, noting that the target spatial operator is a cell-averaged
flux derivative, we prove that the leading truncation error of the MUSCL finite-volume scheme 
is third-order with $\kappa=1/3$. The importance of the diffusion scheme is also discussed: third-order accuracy will be lost when the third-order MUSLC scheme is used with a wrong fourth-order diffusion scheme for convection-diffusion problems. 
Third-order accuracy is verified by thorough numerical experiments for both steady and unsteady problems. This paper is intended to serve as a reference
to clarify confusions about third-order accuracy of the MUSCL scheme, as a guide to correctly analyze and verify the MUSCL scheme for nonlinear equations, and 
eventually as the basis for clarifying third-order unstructured-grid schemes in a subsequent paper. 
\end{abstract}


\section{Introduction}
\label{introduction}

The MUSCL (Monotonic Upstream-centered Scheme for Conservation Laws) approach developed by Van Leer in a series of 
papers \cite{VLeer_Ultimate_IV:JCP1977,VLeer_Ultimate_V:JCP1979}, has become one of the most successful approaches to achieving higher-order accuracy in 
finite-volume methods. MUSCL refers to a comprehensive approach to achieving 
second- and higher-order accuracy in a finite-volume method with a monotonicity preserving mechanism designed for accurately capturing discontinuous solutions. 
In the MUSCL approach, a finite-volume discretization is constructed with cell-averaged solutions stored at cells as numerical solutions, where a higher order accurate flux is achieved at the cell face utilizing higher order accurate solutions reconstructed at the face from left and right cells using these numerical cell-averaged solutions.
For the solution reconstruction, Van Leer proposed a one-parameter-family of solution reconstruction scheme, the $\kappa$-reconstruction scheme, where $\kappa$ is a free parameter, and demonstrated that the finite-volume scheme can achieve third-order accuracy with $\kappa=1/3$ \cite{VLeer_Ultimate_III:JCP1977,VAN_LEER_MUSCL_AERODYNAMIC:J1985}. As one of the most important components of MUSCL being the solution reconstruction step, the
word ``MUSCL" is often used to mean the solution reconstruction. At the same time, ``MUSCL" is also used for the resulting spatial finite-volume discretization: a 
finite-volume scheme constructed by the MUSCL approach is called a MUSCL scheme. 

Despite its popularity over nearly four decades, there still exists some confusion about third-order accuracy of the MUSCL scheme as can be seen even in 
relatively recent papers.
For example, it is claimed in Ref.\cite{Liu:CES2011} that the third-order MUSCL scheme is second-order accurate, which is then cited 
in Ref.\cite{AdvancesIndustrialMixing2015}. In Refs.\cite{SongAmano:2001,Review_TVD:JCP2015},
it is mentioned that an advection scheme based on the solution interpolation corresponding to $\kappa=1/2$ 
is said to be third-order accurate, which implies that the MUSCL scheme with $\kappa=1/3$ is not third-order, without any proof nor numerical verifications.
In Ref.\cite{LeonardMokhtari:IJNMF1990}, the $\kappa=1/3$ scheme is categorized as a second-order method, apparently on the assumption that the numerical solution is a point-valued solution; the MUSCL scheme has cell-averages as numerical solutions \cite{VLeer_Ultimate_IV:JCP1977,VLeer_Ultimate_V:JCP1979}. 
Moreover, Ref.\cite{WuWangSun_NonExistence:1998} claims to have proved that the MUSCL scheme 
cannot be third-order accurate for nonlinear equations.
Later, this reference was cited as a proof of non-existence of third- and higher-order MUSCL schemes \cite{DebiezDerieux:CF2000} and subsequently 
used as the basis for the development of low-dissipation schemes (rather than higher-order) as discussed in Refs.\cite{AbalakinDrvieuxKozubskaya_INRIA_report:2002,CamarriSalvettiKoobusDerieux:CF2004}. Furthemore, Ref.\cite{WuWangSun_NonExistence:1998} is still cited in a very recent reference \cite{AbalakinBakhvalovKozubskaya:IJNMF2015} (2015) 
as a proof of non-existence of third- and higher-order MUSCL schemes. 
In Ref.\cite{AbalakinKozubskayaDervieux:IJA2014}, the same reference \cite{WuWangSun_NonExistence:1998} is cited and it is stated that for nonlinear equations, the
MUSCL scheme is not third-order if the numerical solution is treated as a point value, but a higher-order scheme can be developed based on a direct flux extrapolation.
 The statement is true, but the scheme is no longer considered as a MUSCL scheme if the numerical solution is a point value (it is then 
a conservative finite-difference scheme \cite{Shu_Osher_Efficient_ENO_II_JCP1989}). Therefore, their statement is not true about the MUSCL scheme. 


The truth is that the MUSCL scheme can have arbitrary high-order accuracy. In fact, over the last four decades, numerous high-order finite-volume schemes have been developed based on the MUSCL approach not only for Cartesian grids \cite{FVWENO:JSC2014,FVWENO:AMC2016,TamakiImamura:CF2017} but also for unstructured grids \cite{barth_frederickson_AIAA1990,Luo_etal_rDG_arbitrary:jcp2010,jalali_gooch:CF2017}, just to name a few.  
A major reason for the confusion seems to lie in the fact that the truncation error analysis is not as straightforward for nonlinear equations as for linear equations. To the best of the author's knowledge, truncation error analyses of the MUSCL scheme applied to a nonlinear equation are found 
only in Refs.\cite{Koren:Report1993,Rider:2006} up to a second-order error, and in Ref.\cite{WuWangSun_NonExistence:1998} up to a third-order error. 
Ref.\cite{Koren:Report1993} states that the MUSCL scheme can be third-order accurate with $\kappa=1/3$ for steady linear and nonlinear conservation laws, 
but the MUSCL scheme is actually third-order with $\kappa=1/3$ for a general nonlinear conservation law as we will show in this paper.
More importantly, the proof presented in Ref.\cite{WuWangSun_NonExistence:1998} is incomplete and their conclusion about the non-existence of a third-order MUSCL scheme is false. In proving third-order accuracy, one has to be very careful about the distinction between the point value and the cell average when
expanding the scheme in the Taylor 
series (this distinction is not made in Refs.\cite{WuWangSun_NonExistence:1998,Koren:Report1993}). Another important point missing
in Refs.\cite{WuWangSun_NonExistence:1998,Koren:Report1993} is that the operator approximated by the finite-volume scheme is not the point-valued flux derivative at a cell center but the cell-averaged flux derivative. Missing this point always leads to a second-order truncation error. There have been efforts already in the 1990's to clarify the correct orders of accuracy for various convection schemes \cite{LeonardMokhtari:IJNMF1990,JohnsonMackinnon:CANM1992,ChenFalconer:AWR1994,Leonard_AMM1994,Leonard_AMM1995} and the distinction between the point-valued and cell-averaged operators was recognized (especially by Leonard \cite{LeonardMokhtari:IJNMF1990,Leonard_AMM1994,Leonard_AMM1995}), but analyses were performed only for linear convection equations and the numerical solution seems to have always been considered as a point value. In fact, schemes with point-valued numerical solutions are not MUSCL. As mentioned above (and will be further clarified in a subsequent paper), the MUSCL scheme will become a completely different scheme if used with the numerical solution is taken as a point value.

In this paper, we will provide a careful and detailed truncation error analysis of the MUSCL scheme applied to a nonlinear conservation law, and prove that the MUSCL
scheme can achieve third-order accuracy with $\kappa=1/3$. This paper is intended to serve as a reference
to clarify the confusion about third-order accuracy of the MUSCL scheme and the choice of the diffusion scheme to preserve third-order accuracy, which is an important topic but has rarely discussed in the literature, and also as a guide to analyzing the MUSCL and related schemes for nonlinear equations. In order to minimize the chances of generating 
additional misunderstanding or confusion, we provide a full detail of each step in the truncation error analysis (which could be too much for a standard article) and thorough numerical studies to verify every claim deduced from the analysis and the definition of the MUSCL scheme. 

Finally, we remark that there is a bigger motivation behind this work. Our ultimate goal is to clarify third-order unstructured-grid 
finite-volume schemes used in practical computational fluid dynamics solvers but largely confused in their mechanisms to achieve third- and possibly higher-order
accuracy (e.g., third-order U-MUSCL with $\kappa=1/2$ \cite{burg_umuscl:AIAA2005-4999}, $\kappa=1/3$ \cite{VAN_LEER_MUSCL_AERODYNAMIC:J1985}, 
$\kappa=0$ \cite{katz_work:JCP2015,nishikawa_liu_source_quadrature:jcp2017}). The clarification is important for truly achieving third-order accuracy and taking full advantage of some economical third-order unstructured-grid schemes. In seeking the clarification, we have found that the confusion is largely rooted in the ever-present confusion over third-order convection schemes: the MUSCL scheme and the QUICK scheme \cite{Leonard_QUICK_CMAME1979}. This paper is the first paper in a series. The second paper will focus on the QUICK scheme.  

The paper is organized as follows. In Section \ref{cell_average}, we discuss the integral form of a nonlinear conservation law as a basis for 
a finite-volume discretization and point out the difference between the cell averaged and the point value solutions and operators.
In Section 3, we describe the MUSCL scheme with the $\kappa$-reconstruction scheme applied to a nonlinear conservation law,
and make it clear that the MUSCL scheme is based on cell-averaged solutions.
In Section 4, we derive the truncation error of the MUSCL scheme and prove that third-order accuracy is achieved only with $\kappa=1/3$.
In Section 5, we derive a family of diffusion schemes compatible with the third-order MUSCL scheme.
In Section 6, we present a series of numerical test results to verify the analysis and demonstrate third-order accuracy in the cell-averaged solution, not
in the point-valued solution unless it is accurately recovered from the cell-averaged solution.
In Section 7, we conclude the paper with final remarks.
 
\section{Exact Integral Form with Cell Average}
\label{cell_average}

  \begin{figure}[th!]
    \centering
      \hfill 
          \begin{subfigure}[t]{0.61\textwidth}
        \includegraphics[width=0.99\textwidth]{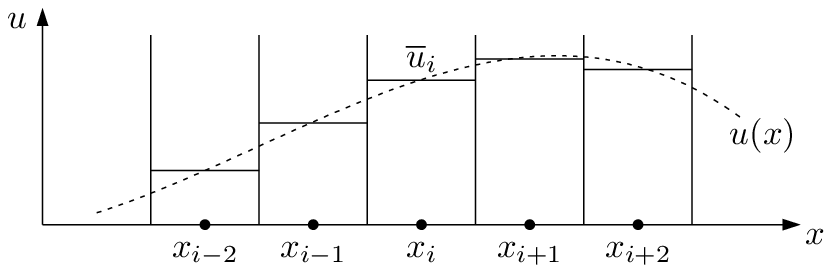} 
            \caption{
\label{fig:oned_fv_data_p0}%
Cell-averaged solutions and control volumes on a uniform grid.
} 
      \end{subfigure}
      \hfill 
          \begin{subfigure}[t]{0.35\textwidth}
              \includegraphics[width=0.99\textwidth]{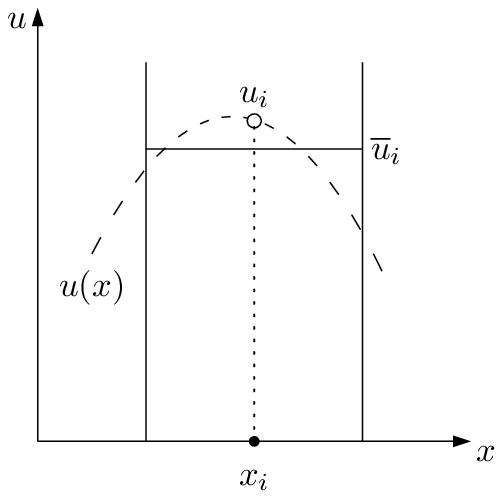} 
            \caption{
\label{fig:oned_fv_data_ca_point}%
The cell-averaged solution and the point-valued solution within the cell $i$. 
}
      \end{subfigure}
      \hfill
            \caption{
\label{fig:oned_data}%
Cell-averaged solutions on a uniform grid and the comparison of the cell average and the point value in a cell $i$.
} 
\end{figure}

Consider a one-dimensional conservation law in the differential form:
\begin{eqnarray}
u_t  + f_x= {s(x)},
\label{diff_form}
\end{eqnarray}
where ${u}$ is a solution variable, $f$ is a flux and ${s}(x)$ is a forcing term, and the subscripts $t$ and $x$ denote the partial derivatives 
with respect to time and space, respectively.
To derive an integral form, we consider a one-dimensional grid with a uniform 
spacing $h$: $x_{i+1} -x_i = h$, $i=1,2,3, \cdots, n$, where $n$ is an integer. See Figure \ref{fig:oned_fv_data_p0}.
In this paper, we focus on the interior of a domain and do not discuss any boundary effects.
Then, we integrate the differential form (\ref{diff_form}) over a control volume around $x=x_i$,
$x \in [x_{i-1/2}, x_{i+1/2} ] =  [x_{i} - h/2, x_{i}+h/2 ] $:
\begin{eqnarray}
\int_{x_i-h/2}^{x_i+h/2} u_t  \,  dx  +  \int_{x_i-h/2}^{x_i+h/2} f_x  \, dx =  \int_{x_i-h/2}^{x_i+h/2}  {s}(x) \, dx,
\end{eqnarray}
and obtain
\begin{eqnarray}
  \frac{d  \overline{u}_i }{dt}     + \frac{1}{h}  [  f(u_{i+1/2}) - f(u_{i-1/2}) ]   = \overline{s}_i,  
\end{eqnarray}
where $\overline{u}$ and $\overline{s}$ are the cell-averaged solution and forcing term:
\begin{eqnarray}
  \overline{u}_i   =   \frac{1}{h}  \int_{x_i-h/2}^{x_i+h/2} {u} \, dx ,\quad
  \overline{s}_i   =   \frac{1}{h} \int_{x_i-h/2}^{x_i+h/2} {s}(x) \, dx , 
\end{eqnarray}
and $u_{i-1/2}$ and $u_{i+1/2}$ are the point-valued solutions at the left and right faces, respectively.

Note that the integral form is exact at this point and no approximation has yet been made. It thus 
represents the exact evolution equation for the cell-averaged solution $ \overline{u}_i$. 
Then, a natural discretization approach is to store the cell-averaged solution at a cell as a numerical solution, 
compute the left and right fluxes with point-valued solutions reconstructed from the cell-averages, and then 
update the cell-averaged solution in time. This is the basis of the finite-volume method and the essence of the MUSCL approach of
Van Leer for achieving second- and higher-order accuracy \cite{VLeer_Ultimate_IV:JCP1977,VLeer_Ultimate_V:JCP1979}.
Note also that it is absolutely clear already at this point from the exactness of the integral form that the MUSCL finite-volume method is arbitrarily high-order accurate with
an arbitrarily high-order of solution reconstruction scheme. Therefore, the claim that the MUSCL scheme cannot be
third-order \cite{WuWangSun_NonExistence:1998} is false. 

However, a confusion can easily arise when one attempts to prove third-order accuracy by a truncation error 
analysis for nonlinear equations. 
One of the main reasons is the lack of distinction of the cell-averaged solution
$\overline{u}_i$ and the point-valued solution $u_i = u(x_i)$, which differ by a second-order error (see Figure \ref{fig:oned_fv_data_ca_point}): 
\begin{eqnarray}
 \overline{u}_{i} = 
 \frac{1}{h} \int_{x_i - h/2}^{x_i+h/2} u(x) \, dx  = u_i + \frac{1}{24} (u_{xx}) h^2 + O(h^4),
 \label{cell_averaged_sol_Taylor}
\end{eqnarray}
where $u(x)$ is a smooth solution that can be expanded around $x=x_i$ as
\begin{eqnarray}
u(x) = u_i + (u_x) (x-x_i)  + \frac{1}{2} (u_{xx}) (x-x_i)^2  + \frac{1}{6} (\partial_{xxx} u) (x-x_i)^3   + \frac{1}{24} (\partial_{xxxx} u) (x-x_i)^4 + O(h^5).
\label{point_taylor_00}
\end{eqnarray}
Another important point that can be easily missed is that it is not the flux derivative $f_x$ but the cell-averaged flux derivative $\overline{f_x}$ that 
the flux difference $[  f(u_{i+1/2}) - f(u_{i-1/2}) ] / h$ represents: 
\begin{eqnarray}
  \frac{d  \overline{u}_i }{dt}  +  \overline{f_x} =   \overline{s}_i, 
  \label{integral_form_operator}
\end{eqnarray}
where
\begin{eqnarray}
\overline{f_x}  =   \frac{1}{h} \int_{x_i - h/2}^{x_i+h/2} f_x \, dx  =  \frac{1}{h}  [  f(u_{i+1/2}) - f(u_{i-1/2}) ] .
\end{eqnarray}
Therefore, the truncation error must be derived for the cell-averaged flux derivative  $\overline{f_x}$, 
not the flux derivative $f_x$ at a point. To clarify the confusion, we provide a detailed derivation of the third-order truncation error of the MUSCL scheme for a general nonlinear conservation law.

\section{Third-Order MUSCL Scheme}
\label{third_order_muscl}


\subsection{Finite-volume discretization}
\label{third_order_muscl_res}

On the one-dimensional grid defined in the previous section, we store the cell-averaged solution $\overline{u}_i$ at a cell $i$ as a numerical solution, and 
define a finite-volume scheme as the direct application of the integral form:
\begin{eqnarray}
  \frac{d  \overline{u}_i }{dt}  +   \frac{1}{h}  [ F(u_{i+1/2,L},u_{i+1/2,R} ) - F(u_{i-1/2,L}, u_{i-1/2,R}) ]   =  \overline{s}_i,
\label{fv_exact_form}
\end{eqnarray}
where $F$ denotes a numerical flux computed from two point-valued solutions reconstructed at a face, e.g., $u_{i+1/2,L}$ and $u_{i+1/2,R}$ 
at the right face. In this paper, we consider an upwind numerical flux in the form:
\begin{eqnarray}
F(u_L,u_R)  = \frac{1}{2} \left[  f(u_L) + f(u_R)  \right]  - \frac{1}{2} D ( u_R - u_L),
\label{upwnd_flux}
\end{eqnarray}
where $D$ is a dissipation coefficient $D= | \partial f / \partial u|$,
and $u_L$ and $u_R$ are reconstructed solutions at a face from the left and right cells, respectively. See 
Figure \ref{fig:oned_fv_data_quad} for an example of piecewise quadratic polynomials reconstructed from the cell-averaged solutions.
Here, we assume that the forcing term will be integrated exactly or by a sufficiently accurate quadrature formula. Then, the above discretization is exact 
if the reconstructed solutions are computed exactly; of course, they cannot be exact, and thus the solution reconstruction accuracy
determines the order of accuracy of the finite-volume discretization.
 
\subsection{Reconstruction scheme}
\label{third_order_muscl_reconstruct}

  \begin{figure}[t]
    \centering 
        \includegraphics[width=0.7\textwidth]{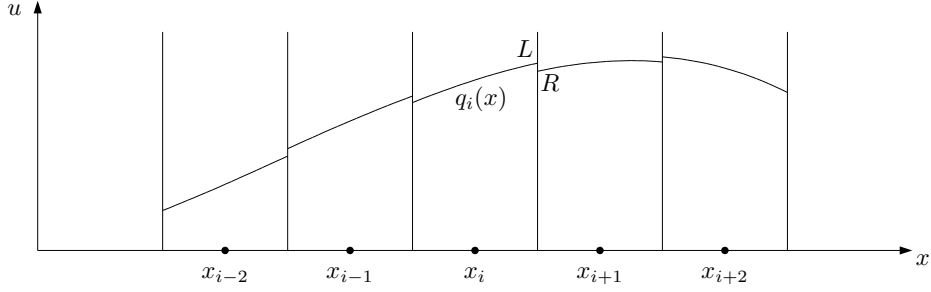} 
            \caption{
\label{fig:oned_fv_data_quad}%
Quadratically reconstructed point-valued solution within each control volume on a uniform grid.
} 
\end{figure}

For the solution reconstruction, we consider Van Leer's $\kappa$-reconstruction scheme, which
first appeared in Ref.\cite{VLeer_Ultimate_III:JCP1977} and later was formulated specifically for a finite-volume discretization in Ref.\cite{VAN_LEER_MUSCL_AERODYNAMIC:J1985}: 
for the right face at $i+1/2$ (see Figure \ref{fig:oned_fv_data_quad}), 
\begin{eqnarray}
u_L &=&\overline{u}_{i}  +  (u_x)_{i} \left(  \frac{h}{2} \right) +  \frac{3 \kappa}{2}  (u_{xx})_{i} \left[   \left( \frac{h}{2} \right)^2 -   \frac{h^2}{12} \right], 
\label{kappa_uL} \\ [2ex]
u_R &=& \overline{u}_{i+1} +  (u_x)_{i+1} \left(-  \frac{h}{2} \right) +  \frac{3 \kappa}{2}(u_{xx})_{i+1} \left[  \left(  -\frac{h}{2} \right)^2 -   \frac{h^2}{12} \right], 
\label{kappa_uR}
\end{eqnarray}
where
\begin{eqnarray} 
 (u_x)_{i}  = \frac{  \overline{u}_{i+1} -  \overline{u}_{i-1}  }{2h}, \quad 
 (u_{xx})_{i}  =  \frac{  \overline{u}_{i+1}  -2 \overline{u}_{i}+  \overline{u}_{i-1}  }{h^2}, 
 \label{fd_aprpox_at_i}
\end{eqnarray}
\begin{eqnarray} 
 (u_x)_{i+1}  = \frac{  \overline{u}_{i+2} -  \overline{u}_{i}  }{2h}, \quad 
 (u_{xx})_{i+1}  =  \frac{  \overline{u}_{i+2}  -2 \overline{u}_{i+1}+  \overline{u}_{i}  }{h^2},
 \label{fd_aprpox_at_ip1}
\end{eqnarray}
or they can be simplified as
\begin{eqnarray}
u_L &=& \frac{1}{2}  \left( \overline{u}_{i} +   \overline{u}_{i+1}   \right)- \frac{1-\kappa}{4} \left(  \overline{u}_{i+1}  -2 \overline{u}_{i}+  \overline{u}_{i-1} \right) ,  
\label{kappa_uL_simple} \\ [2ex]
u_R &=& \frac{1}{2}  \left( \overline{u}_{i+1} +   \overline{u}_{i}   \right)- \frac{1-\kappa}{4} \left(  \overline{u}_{i+2}  -2 \overline{u}_{i+1}+  \overline{u}_{i} \right).
\label{kappa_uR_simple}
\end{eqnarray}
Note that we have denoted $u_{i-1/2,L}$ and $u_{i-1/2,R}$ simply as $u_L$ and $u_R$ for brevity; we consider only the right face across cells $i$ and $i+1$ from 
now on since the left face is treated in a similar manner by shifting the index. Notice that the $\kappa$-reconstruction scheme has been generated from a local quadratic polynomial 
defined in each cell. For example, in the cell $i$ shown in Figure \ref{fig:oned_fv_data_quad}, the quadratic polynomial $q_i(x)$ is given by
\begin{eqnarray}
q_i(x) = \overline{u}_{i} + (u_x)_{i}  (x - x_{i} ) + \frac{1}{2} (u_{xx})_{i} \left[  (x - x_{i} )^2 -   \frac{h^2}{12} \right],
\label{quadratic_poly_iw}
\end{eqnarray}
where
\begin{eqnarray} 
 (u_x)_{i}  = \frac{  \overline{u}_{i+1} -  \overline{u}_{i-1}  }{2h}, \quad 
 (u_{xx})_{i}  =  \frac{  \overline{u}_{i+1}  -2 \overline{u}_{i}+  \overline{u}_{i-1}  }{h^2},
\end{eqnarray}
which gives $u_L$ at $x=x_i+h/2$ as in Equation (\ref{kappa_uL}).
It is important to note that this quadratic polynomial is a reconstruction of a point-valued solution from cell averages, which is clear from the fact that its cell-average is $\overline{u}_{i}$:
\begin{eqnarray}
\frac{1}{h} \int_{x_i - h/2}^{x_i+h/2} q_i(x) \, dx =  \overline{u}_{i}.
\end{eqnarray}
Note also that the derivatives $ (u_x)_{i}$ and $ (u_{xx})_{i}$ are point values as can be easily proved as
\begin{eqnarray}
\left. \frac{d q_i(x) }{d x } \right|_{x=x_i} =   (u_x)_{i}, \quad
\left. \frac{d^2 q_i(x) }{d x^2 } \right|_{x=x_i} =   (u_{xx})_{i}.
\end{eqnarray}
As pointed out in Ref.\cite{VAN_LEER_MUSCL_AERODYNAMIC:J1985}, the $\kappa$-reconstruction scheme corresponds to Fromm's scheme with $\kappa=0$, a quadratic point-valued solution reconstruction (from cell averages) with $\kappa=1/3$, a quadratic interpolation scheme (i.e., the QUICK interpolation scheme \cite{Leonard_QUICK_CMAME1979}) with $\kappa=1/2$, and the central 
scheme with $\kappa=1$. As we will show, third-order accuracy is achieved for the finite-volume scheme only with $\kappa=1/3$.
 \newline
 \newline
\noindent {\bf Remark}: Readers who are not familiar with the $\kappa$-reconstruction scheme written as in Equations (\ref{kappa_uL}) and (\ref{kappa_uR}) may find the following
list of equivalent forms useful: Equations (\ref{kappa_uL}) and (\ref{kappa_uR}) are equivalent to 
\begin{eqnarray}
u_L  = \overline{u}_{i}  +  \frac{1}{4} \left\{
   (1-\kappa) \Delta^-_i 
+ (1+\kappa) \Delta^+_i 
\right\} ,  \label{kappa_uL_useful_form} 
\quad
u_R  =
\overline{u}_{i+1}  -  \frac{1}{4} \left\{
   (1-\kappa) \Delta^+_{i+1} 
+ (1+\kappa) \Delta^-_{i+1} 
\right\} ,
\label{kappa_scheme_diff_form}
\end{eqnarray}
where
\begin{eqnarray}
 \Delta^-_i   =  \overline{u}_{i}   -  \overline{u}_{i-1} ,
 \quad
  \Delta^+_i  =  \overline{u}_{i+1}  - \overline{u}_{i},
  \label{kappa_diff_plus_minus}
\end{eqnarray}
which may be a convenient form to incorporate classical limiters \cite{johnlynn:phd,Hirsch_VOL2}, or 
\begin{eqnarray}
u_L = \overline{u}_{i}     + \frac{\kappa}{2} ( \overline{u}_{i+1} - \overline{u}_{i}  ) +  (1-\kappa ) (u_x)_{i}    \frac{h}{2}, \quad
u_R = \overline{u}_{i+1}  -   \frac{\kappa}{2} ( \overline{u}_{i+1} - \overline{u}_{i}  )  -  (1-\kappa ) (u_x)_{i+1} \frac{h}{2}, 
\label{umuscl_general_form01}
\end{eqnarray}
which has been found useful for extending $\kappa$-reconstruction scheme to unstructured grids \cite{DebiezDerieux:CF2000,burg_umuscl:AIAA2005-4999}, or 
\begin{eqnarray}
{u}_L = \kappa  \frac{\overline{u}_{i} + \overline{u}_{i+1}}{2} +   (1-\kappa) \left[ \overline{u}_j  + \frac{1}{2}(u_x)_{i}  h \right] , \quad
{u}_R = \kappa  \frac{\overline{u}_{i+1} + \overline{u}_{i}}{2} +   (1-\kappa) \left[ \overline{u}_{i+1} - \frac{1}{2}(u_x)_{i+1}  h \right].
 \label{umuscl_L2}
\end{eqnarray}
which has been found useful for explaining how an unstructured-grid version of $\kappa$-reconstruction scheme can lose the linear exactness on unstructured grids \cite{nishikawa_LP_UMUSCL:JCP2020}. Note also that the reconstruction scheme proposed in Ref.\cite{DebiezDerieux:CF2000}, which they call the $\beta$-scheme with a parameter $\beta$, is equivalent to the $\kappa$-reconstruction scheme with $\kappa=1-2\beta$; thus it is a re-arrangement of the $\kappa$-reconstruction scheme. The scheme proposed in Ref.\cite{yang_harris:AIAAJ2016} is not equivalent but closely related to the $\kappa$-reconstruction scheme. With an additional term added, it is claimed to achieve third-order accuracy with $\kappa=-1/6$ for point-valued numerical solutions. 

\section{Accuracy of Third-Order MUSCL Scheme}
\label{third_order_muscl_accuracy}


To prove third-order accuracy of the MUSCL scheme with $\kappa=1/3$, we first derive the so-called modified equation for the MUSCL scheme by expanding the numerical solution
in a Taylor series \cite{Hirt_ModifiedEq:JCP1968}. Then, we identify the truncation error as the leading term in the difference between the modified equation and
the target conservation law. For convenience, we will use the Taylor expansion of a point-valued solution and then take its cell-average to expand 
the cell-averaged 
solution. 

In the rest of the section, we proceed as follows. First, we examine the accuracy of the solution reconstruction and show that the averaged
solution at a face is exact for a cubic solution with $\kappa=1/3$. Then, we prove that the same is true for the averaged flux term in the 
numerical flux (\ref{upwnd_flux}); this is an important point for nonlinear equations. Furthermore, we will prove that the dissipation term in the 
numerical flux (\ref{upwnd_flux}) is a fourth-order quantity at a face and therefore will only generate a third-order truncation error.
Finally, we derive a modified equation of the MUSCL scheme by converting the resulting point-valued flux derivative to a cell-averaged derivative.
Below, we will present as much detail as possible to leave no room for misunderstanding.

\subsection{Accuracy of reconstruction}
\label{third_order_muscl_reconstruct_acc}

To examine the accuracy of the $\kappa$-reconstruction scheme, we consider a Taylor expansion of a smooth point-valued solution $u(x)$ around a cell center $x=x_i$:
\begin{eqnarray}
u(x) = u_i + (u_x) (x-x_i)  + \frac{1}{2} (u_{xx}) (x-x_i)^2  + \frac{1}{6} (u_{xxx}) (x-x_i)^3   + \frac{1}{24}   (u_{xxxx}) (x-x_i)^4 + O(h^5),
\label{point_taylor}
\end{eqnarray}
where $u_i$ is a point value at $x=x_i$. It is very important to note here that the derivatives such as $(u_x)$ and $(u_{xx})$ are also point values 
defined at $x=x_i$, which can be easily verified by differentiating $u(x) $ and evaluating the result at $x=x_i$. Here, we do not use the subscript $i$ for the derivatives 
to distinguish them from the finite-difference approximations in Equations (\ref{fd_aprpox_at_i}) and (\ref{fd_aprpox_at_ip1}). It is possible and actually more convenient for our purpose to use a Taylor expansion of a cell-averaged solution with cell-averaged derivatives \cite{TamakiImamura:CF2017}, but we do not employ such an approach here to illustrate and emphasize the importance of recognizing the difference between the cell average and the point value. 

The exact reconstruction of a smooth solution at the right face is given by  
\begin{eqnarray}
u_{i+1/2}^{exact} = u(x_i+h/2) =  {u}_{i}  + \frac{1}{2} (u_x) h  + \frac{1}{8} (u_{xx}) h^2   + \frac{1}{48} (u_{xxx}) h^3  + \frac{1}{384}   (u_{xxxx})  h^4 + O(h^5).
\label{exact_reconstruction_uL}
\end{eqnarray}
This is the target expression we wish to approximate as accurately as possible with the $\kappa$-reconstruction scheme.
The $\kappa$-reconstruction scheme is expected to be exact up to the quadratic term since it is 
derived from a quadratic polynomial as discussed in the previous section, which is indeed true. To see this, 
we express (\ref{exact_reconstruction_uL}) in terms of the cell-average stored at the cell $i$: take a cell-average of Equation (\ref{point_taylor}), 
\begin{eqnarray}
 \overline{u}_{i} = 
 \frac{1}{h} \int_{x_i - h/2}^{x_i+h/2} u(x) \, dx  = u_i + \frac{1}{24} (u_{xx}) h^2  + \frac{1}{1920}   (u_{xxxx}) h^4  + O(h^6),
\end{eqnarray}
solve it for $u_i$, and substitute it into Equation (\ref{point_taylor}) to get
\begin{eqnarray}
u(x) = \overline{u}_{i}  + (u_x) (x-x_i)  + \frac{1}{2} (u_{xx}) \left[ (x-x_i)^2 - \frac{h^2}{12}  \right] + \frac{1}{6}  (u_{xxx})(x-x_i)^3 + O(h^4), 
\end{eqnarray}
which gives
 \begin{eqnarray}
u_{i+1/2}^{exact} &=&  \overline{u}_{i}  + (u_x) \frac{h}{2}  + \frac{1}{2} (u_{xx}) \left[ \left( \frac{h}{2} \right)^2 - \frac{h^2}{12}  \right] + \frac{1}{6}  (u_{xxx})  \left( \frac{h}{2} \right)^3 + O(h^4)  \\ [2ex]
&=& \overline{u}_{i}  + \frac{1}{2} (u_x) h  + \frac{1}{12} (u_{xx}) h^2   + \frac{1}{48}   (u_{xxx}) h^3 + O(h^4).
\end{eqnarray}
Comparing the $\kappa$-reconstruction scheme (\ref{kappa_uL}) with this exact expression, we find $\kappa=1/3$ does produce the exact 
reconstruction up to the quadratic term. On uniform, however, it is actually exact up to the cubic term when $u_L$ and $u_R$ are averaged. 
This fact is pointed out in Ref.\cite{WatersonDeconinck_Convection:JCP2007} by quoting Ref.\cite{agarwal_aiaa1981-0112}, where a third-order
finite-difference scheme is proposed, which is equivalent to the MUSCL scheme with $\kappa=1/3$ for a linear advection equation (not for nonlinear equations). 
\newline
\newline
\noindent {\bf Remark}: The cubic exactness is a special property of a quadratically exact algorithm on a regular grid; the quadratic exactness is sufficient to design a third-order scheme on arbitrary grids. The resulting one-order higher truncation error is the reason that the truncation error order matches the discretization error (i.e., solution error) order on regular grids; the truncation error order is typically one-order lower on irregular grids. 
See Refs.\cite{Boris_Jim_NIA2007-08,DiskinThomas:ANM2010,Katz_Sankaran_JCP:2011,nishikawa_liu_source_quadrature:jcp2017} and references therein for further details about accuracy on irregular grids. 
\newline
\newline
To prove the cubic exactness, we expand the $\kappa$-reconstruction schemes (\ref{kappa_uL}) and (\ref{kappa_uR}) by substituting the expansions of the cell-averages: 
 \begin{eqnarray}
\overline{u}_{i-1} &=&  \frac{1}{h} \int_{x_i - 3h/2}^{x_i-h/2} u(x) \, dx  = u_i  -  (\partial_{x} u) h + \frac{13}{24} (u_{xx}) h^2  - \frac{5}{24} (u_{xxx}) h^3  + O(h^4), 
\label{expanded_ca_im1} \\ [2ex]
\overline{u}_{i }   &=&  \frac{1}{h} \int_{x_i - h/2}^{x_i+h/2} u(x) \, dx  = u_i + \frac{1}{24} (u_{xx}) h^2  + \frac{1}{1920} (\partial_{xxxx} u) h^4  + O(h^6), \\ [2ex]
\overline{u}_{i+1} &=&  \frac{1}{h} \int_{x_i + h/2}^{x_i+3h/2} u(x) \, dx  = u_i  +  (\partial_{x} u) h + \frac{13}{24} (u_{xx}) h^2  + \frac{5}{24} (u_{xxx}) h^3  + O(h^4), \\ [2ex]
\overline{u}_{i+2} &=&  \frac{1}{h} \int_{x_i + 3h/2}^{x_i+ 5h/2} u(x) \, dx  =u_i  + 2 (\partial_{x} u) h + \frac{49}{24} (u_{xx}) h^2  +\frac{17}{12}  (u_{xxx})h^3  + O(h^4) ,
\label{expanded_ca_ip2}
\end{eqnarray}
and obtain 
 \begin{eqnarray}
u_L &=& u_i  + \frac{1}{2} (u_x) h  + \frac{1}{4}  (u_{xx}) \left(   \kappa + \frac{1}{6}   \right) h^2 + \frac{5}{48}  (u_{xxxx})   h^3
+  \frac{60 \kappa+1}{1920}   (u_{xxxxx})  h^4
  + O(h^5), \nonumber \\ [2ex]
  u_R &=& u_i  + \frac{1}{2} (u_x) h  + \frac{1}{4}  (u_{xx}) \left(   \kappa + \frac{1}{6}   \right) h^2 - \frac{1}{8} \left(   \frac{7}{6}  - 2 \kappa \right)  (u_{xxxx})  h^3
 -  \left(   \frac{239}{1920}  - \frac{5 \kappa}{32}  \right) (u_{xxxxx})  h^4
  + O(h^5), \nonumber 
\end{eqnarray}
thus
 \begin{eqnarray}
 \frac{u_L + u_R}{2} = u_i  + \frac{1}{2} (u_x) h  + \frac{1}{4}  (u_{xx}) \left(   \kappa + \frac{1}{6}   \right) h^2 - \frac{1}{8} \left(   \frac{1}{6}  - \kappa \right)  (u_{xxxx})  h^3
 - \frac{1}{16} \left(   \frac{119}{120}  - \frac{3 \kappa}{2}  \right) (u_{xxxxx})   h^4
  + O(h^5),
\end{eqnarray}
which matches the exact reconstruction ({\ref{exact_reconstruction_uL}) up to the cubic term if 
\begin{eqnarray}
   \kappa + \frac{1}{6}    = \frac{1}{2},  \quad    \frac{1}{6}  - \kappa = - \frac{1}{6},
\end{eqnarray}
both of which are satisfied with $\kappa=1/3$, thus resulting in
 \begin{eqnarray}
 \frac{u_L + u_R}{2} = u_i  + \frac{1}{2} (u_x) h  + \frac{1}{8}  (u_{xx})   h^2 + \frac{1}{48}  (u_{xxxx})   h^3 - \frac{59}{1920}    (u_{xxxxx})  h^4
  + O(h^5).
  \label{expanded_sol_average}
\end{eqnarray}
Therefore, the $\kappa$-reconstruction scheme leads to the average of the reconstructed solutions exact for a cubic functions on a uniform grid. 
For a linear problem with $f=a u$, where $a$ is a constant, this means that the flux is reconstructed exactly for a cubic flux and thus it leads to 
a third-order MUSCL scheme for linear equations. However, for nonlinear equations, we must prove that the average of the nonlinear fluxes evaluated with the
reconstructed solutions $ [ f(u_L)+ f(u_R) ]/2 $ is also exact for a cubic flux. It requires a careful derivation as we will discuss in the next section.


\subsection{Accuracy of flux}
\label{third_order_muscl_flux_acc}

For nonlinear equations, the flux is a nonlinear function of $u$ and thus it is not immediately clear if the averaged flux 
$ [ f(u_L)+ f(u_R) ]/2 $ is also exact for a cubic flux. However, it is indeed exact for a cubic flux.
To prove this, we will expand the averaged flux, but the expansion needs to be performed very carefully. 
Consider $f(u_L)$: 
\begin{eqnarray}
f(u_L)   = f \left(  \overline{u}_{i}  +  (u_x)_{i} \left(  \frac{h}{2} \right) +  \frac{3 \kappa}{2}  (u_{xx})_{i} \left[   \left( \frac{h}{2} \right)^2 -   \frac{h^2}{12} \right] \right), 
\end{eqnarray}
which one might simply expand with respect to $\overline{u}_{i}$,
\begin{eqnarray}
f(u_L)   = f(\overline{u}_{i})  + O(( u_L -  \overline{u}_{i})^2 ).
\end{eqnarray}
However, we need a point-wise expansion of the flux around the cell center based on $f({u}_{i})$, not $f(\overline{u}_{i})$, which differs 
by $O(h^2)$ and does not make sense. This is an important point because the flux derivatives arising from the flux expansion and to be used to 
represent the target equation in the modified equation should be point values, not those evaluated by the cell-averaged solution.
The distinction is not made and it seems to be one of the major problems in the proof presented in Ref.\cite{Liu:CES2011}.
To derive a correct truncation error, we write
\begin{eqnarray}
f(u_L)   = f(u_i +  du)   , \quad du_L = u_L-u_i,
\end{eqnarray}
and expand it around the point value $f({u}_{i})$ as
\begin{eqnarray}
f(u_L)   = f(u_i)   + (f_u) du_L + \frac{1}{2} (f_{uu}  ) du_L^2  + \frac{1}{6} ( f_{uuu} ) du_L^3 + O(du^4),
\end{eqnarray}
and similarly for $f(u_R)$, 
\begin{eqnarray}
f(u_R)   = f(u_i)   + (f_{u}) du_R + \frac{1}{2} ( f_{uu}) du_R^2  + \frac{1}{6} ( f_{uuu}) du_R^3 + O(du^4),
\end{eqnarray}
where  $du_R= u_R-u_i$. Note that all the derivatives, e.g., $(f_{u})$ and $(f_{uu}) $, are point values at
the cell center $x=x_i$.
 Then, we take the average to get 
 \begin{eqnarray}
 \frac{f(u_L) + f(u_R)  }{2} &=&
  f(u_i)   + \frac{1}{2} ( f_{u} ) (u_x) h  + \frac{1}{8} \left[  \frac{6 \kappa+1}{3}  ( f_{u} ) (u_{xx})  + (f_{uu})(u_x)^2   \right] h^2 \nonumber \\
  &+&  \frac{1}{48} \left[         ( f_{uuu} )(u_x)^3 +  \left( 6  \kappa +1  \right) (f_{uu} ) (\partial_{x} u)  (u_{xx})   
  +  \left( 6 \kappa -1 \right) ( f_{u} )   (\partial_{xxx} u)   
                        \right] h^3  
  + O(h^4).
  \label{ave_flux_at_right_face_kappa}
\end{eqnarray}
Comparing it with the exact flux reconstruction expressed in the Taylor series,
\begin{eqnarray}
f_{j+1/2}^{exact} &=&  f({u}_{i})  + \frac{1}{2} (f_x) h  + \frac{1}{8} (f_{xx}) h^2   + \frac{1}{48}  (f_{xxx}) h^3   + O(h^4), \nonumber \\ [2ex]
&=& f({u}_{i})  +   \frac{1}{2} ( f_{uu} )(u_x) h  +  \frac{1}{8} \left[    (f_{u} ) (u_{xx})  + (f_{uu})(u_x)^2   \right] h^2 \nonumber \\
  &+&  \frac{1}{48} \left[       ( f_{uuu} )(u_x)^3 + 3  (f_{uu} ) ( u_x)  (u_{xx})    +  ( f_{u} )   (u_{xxx})   
                        \right] h^3    + O(h^4),
\label{exact_reconstruction_fL}
\end{eqnarray}
we find that the averaged flux matches the exact flux up to the cubic term if 
 \begin{eqnarray}
 \frac{6 \kappa+1}{3} = 1, \quad 6  \kappa +1  = 3, \quad  6 \kappa -1 = 1, 
\end{eqnarray}
which are all satisfied with $\kappa=1/3$, thus giving 
 \begin{eqnarray}
 \frac{f(u_L) + f(u_R)  }{2} &=& f({u}_{i})  +   \frac{1}{2} (f_{u})(u_x) h  +  \frac{1}{8} \left[    (f_{u}) (u_{xx})  + (f_{uu} )(u_x)^2   \right] h^2 \nonumber \\
  &+&  \frac{1}{48} \left[       ( f_{uuu} )(u_x)^3 + 3  ( f_{uu} ) ( u_x)  (u_{xx})    +  ( f_{u} )   (u_{xxx})   
                        \right] h^3    + O(h^4).
\end{eqnarray}
Therefore, the average of the left and right fluxes evaluated with the solutions reconstructed with the $\kappa$-reconstruction scheme 
is exact for a cubic flux with $\kappa=1/3$.
 
\subsection{Order of dissipation}
\label{third_order_muscl_dissipation}

In this section, we show that the dissipation term is of $O(h^3)$ and does not contribute to the second-order truncation error. 
Consider the solution jump at the right face, which can be expanded by substituting the expanded cell-averages (\ref{expanded_ca_im1})-(\ref{expanded_ca_ip2}): 
\begin{eqnarray}
u_R - u_L =   (\kappa-1) \left[  \frac{1}{4} (u_{xxx}) h^3 +   \frac{1}{8} (u_{xxxxx}) h^4  + O(h^5)  \right].
\end{eqnarray}
Observe that the jump vanishes for $\kappa=1$ and it is consistent with the fact that $\kappa=1$ corresponds to the central scheme, which 
has no dissipation. For a linear equation, the dissipation coefficient $D$ in  the upwind flux (\ref{upwnd_flux}) is a global constant 
and the leading third-order term will cancel with the same term arising from the expansion of the jump from the left face. 
However, for nonlinear equations, the dissipation coefficient
is generally a function of the solution. It is typically evaluated with the averaged solution: 
\begin{eqnarray}
D = D \left(  \frac{ u_L + u_R  }{2} \right).
\label{D_before_exp}
\end{eqnarray}
Note that other averages can be used but all differ from the arithmetic average by $O(u_R - u_L)=O(h^3)$, which is negligibly small for the purpose of our analysis. 
Hence, it suffices to consider the arithmetic average as in the above equation.
This coefficient needs to be expanded as well in the truncation error analysis. In doing so, we assume that the 
dissipation coefficient is differentiable with a smoothing technique (e.g., see Ref.\cite{Harten_Hyman_JCP_1983}) applied to the absolute value of the eigenvalue
in constructing the absolute Jacobian $D= |\partial f / \partial u|$. Then, we expand the dissipation coefficient (\ref{D_before_exp}) 
by using Equation (\ref{expanded_sol_average}), and find
\begin{eqnarray}
D = D (u_i) + \frac{1}{2} D_{u} (u_x) h + O(h^2).
\end{eqnarray}
where $D_u$ is the derivative of $D$ with respect to $u$ at $x=x_i$. Therefore, the dissipation term at the right face is expanded as
\begin{eqnarray}
D ( u_R - u_L ) &=& 
\left[  D (u_i) + \frac{1}{2} D_{u} (u_x) h + O(h^2) \right] 
 (\kappa-1) \left[  \frac{1}{4} (u_{xxx}) h^3 +   \frac{1}{8} (u_{xxxxx}) h^4  + O(h^5)  \right]  \\[1.2ex]
 &=& 
 D (u_i) \frac{\kappa-1}{4} (u_{xxx}) h^3  + \frac{\kappa-1}{8} \left[   D_u (u_{x}) (u_{xxx}) +(u_{xxxx})    \right]  h^4  + O(h^5) .
\end{eqnarray}
By performing a similar expansion for the dissipation term at the left face, we obtain exactly the same third-order leading term 
(and the same fourth-order term with the opposite sign). 
Therefore, the third-order contribution from the dissipation term will cancel out in the flux difference in Equation (\ref{fv_exact_form}).
Then, as the flux difference is divided by $h$, the leading error coming from the dissipation term will be $O(h^3)$. 
Hence, the truncation error of the dissipation in the MUSCL scheme is already sufficiently small for third-order accuracy.
For this reason, we do not consider the dissipation term in the analysis below.

\subsection{Accuracy of finite-volume scheme}
\label{third_order_muscl_fvreconstruct_acc}

We now derive the spatial truncation error of the MUSCL scheme by substituting the Taylor expansions of 
the fluxes and solutions as derived in the previous sections. Substituting the expansion of the averaged flux (\ref{ave_flux_at_right_face_kappa}) 
at the right face and a similar expression for that at the left face, we obtain
\begin{eqnarray}
 \frac{d  \overline{u}_i }{dt}  + 
 f_x +  \frac{1}{24} \left[     f_{uuu} (u_x)^3 +  \left( 6  \kappa +1  \right) f_{uu} u_x  u_{xx}   +  \left( 6 \kappa -1 \right) f_u  u_{xxx}
\right] h^2
 + O(h^3) 
=   \overline{s}_i,
\label{fv_exact_form_te_not_yet}
\end{eqnarray}
where we have used $f_u u_x = f_x$, which is true since all derivatives are point values at $x=x_i$. 
There is a second-order error term, but the derivation is not completed yet. As mentioned earlier and repeatedly pointed out by Leonard \cite{LeonardMokhtari:IJNMF1990,Leonard_AMM1995}, the spatial operator that the finite-volume discretization is trying to approximate
 is not the pointwise $f_x$ but the cell average $\overline{f_x}$. 
 To derive a correct truncation error, we replace $f_x$ by $\overline{f_x}$. Consider the cell-average of $f_x$.
\begin{eqnarray}
\overline{f_x} =   \frac{1}{h} \int_{x_i - h/2}^{x_i+h/2} f_x \, dx  = f_x + \frac{1}{24} f_{xx} h^2  + \frac{1}{1920}  f_{xxxxx} h^4  + O(h^6),  
\end{eqnarray}
from which we find
\begin{eqnarray}
 f_x  =   \overline{f_x}-  \frac{1}{24} f_{xxx} h^2  -  \frac{1}{1920}  f_{xxxxx} h^4  - O(h^6), 
\end{eqnarray}
and substituting this into Equation (\ref{fv_exact_form_te_not_yet}), we obtaiin
\begin{eqnarray}
 \frac{d  \overline{u}_i }{dt}  + 
 \overline{f_x} +  \frac{1}{24} \left[     f_{uuu} (u_x)^3 +  \left( 6  \kappa +1  \right) f_{uu} u_x  u_{xx}   +  \left( 6 \kappa -1 \right) f_u  u_{xxx} - 
 f_{xxx} 
\right] h^2
 + O(h^3) 
=   \overline{s}_i,
\label{fv_exact_form_te_not_yet000}
\end{eqnarray}
which becomes by $f_{xxx} =  f_{uuu} (u_x)^3 +  3 f_{uu} u_x  u_{xx}   +   f_u  u_{xxx}$
\begin{eqnarray}
 \frac{d  \overline{u}_i }{dt}  + 
 \overline{f_x} +  \frac{3  \kappa -1}{12} \left[    f_{uu} u_x  u_{xx}   +  f_u  u_{xxx} 
\right] h^2
 + O(h^3) 
=   \overline{s}_i.
\label{fv_exact_form_te_not_yet001}
\end{eqnarray}
Then, the second-order term vanishes for $\kappa=1/3$: 
\begin{eqnarray}
 \frac{d  \overline{u}_i }{dt}  + 
 \overline{f_x} 
 + O(h^3) 
=   \overline{s}_i.
\label{fv_exact_form_te_not_yet010}
\end{eqnarray}
This is the correct modified equation for the third-order MUSCL scheme. By comparing with 
the target integral form (\ref{integral_form_operator}), we find that the truncation error is $O(h^3)$. 
Therefore, the MUSCL scheme is third-order accurate with $\kappa=1/3$ for a general nonlinear conservation law. 

Note that we have just proved that the MUSCL scheme cannot be third-order with $\kappa=1/2$. 
However, it does not disprove third-order accuracy of the QUICK scheme 
of Leonard \cite{Leonard_QUICK_CMAME1979} because the QUICK scheme 
is not equivalent to the $k=1/2$ MUSCL scheme. It is still based on the integral form, but the numerical
solution is taken as a point value, in which case the  $\kappa=1/3$ MUSCL scheme is not third-order accurate as analyzed in Ref.\cite{LeonardMokhtari:IJNMF1990}.
In a subsequent paper, we will provide a detailed discussion on the third-order QUICK scheme. 

\section{Fourth-Order MUSCL Diffusion Scheme}
\label{MUSCL_diffusion}

Before we move on to numerical experiments, we consider a MUSCL scheme for diffusion, which 
is required to solve a convection-diffusion equation:
\begin{eqnarray}
f_x - \nu u_{xx} = 0,
\label{conv_diff_eq}
\end{eqnarray}
where $\nu$ is a positive constant. This is an important subject but rarely discussed in the literature. 
To show its significance, we consider the following fourth-order diffusion scheme:
 \begin{eqnarray}
f_x - \nu \frac{  -  \overline{u}_{i-2} + 12 \overline{u}_{i-1} - 22 \overline{u}_i + 12 \overline{u}_{i+1} -  \overline{u}_{i+2}  }{12 h^2} = 0.
  \label{diffusion_scheme_wrong}
\end{eqnarray}
which is indeed fourth-order accurate as a finite-difference scheme,
\begin{eqnarray}
 f_x  - \nu  u_{xx}  +  \frac{37 u_{xxxxx}}{1920} h^4 + O(h^6) = 0,
\end{eqnarray}
but only second-order accurate as a finite-volume scheme approximating the integral form,
\begin{eqnarray}
 \frac{1}{h} \int_{x_i - h/2}^{x_i+h/2} f_x \, dx  -   \frac{1}{h} \int_{x_i - h/2}^{x_i+h/2} (  \nu  u_{xx}) \, dx  +      \frac{\nu}{24} (    u_{xxxx} ) h^2  +  O(h^4) = 0.
\end{eqnarray}
Therefore, if the third-order MUSCL scheme is used with this fourth-order diffusion scheme, the resulting scheme will be
second-order accurate unless the diffusion term is negligibly small. 

To construct a third-order MUSCL scheme for the convection-diffusion equation, we must construct a fourth-order finite-volume diffusion scheme.
Consider the finite-volume discretization of the integral form of the diffusion term:
\begin{eqnarray}
-   \frac{1}{h} \int_{x_i - h/2}^{x_i+h/2} (  \nu  u_{xx}) \, dx  \approx 
  - \frac{1}{h}  \left[  F^d_{i+1/2} -F^d_{i-1/2}  \right] ,
\label{fv__diffusion}
\end{eqnarray}
where $F^{d}$ is the alpha-damping diffusive flux \cite{nishikawa:AIAA2010,nishikawa_general_principle:CF2011},
\begin{eqnarray}
F^d(u_L,u_R)  = -  \frac{1}{2} \left[   \nu (u_x)_L   +  \nu (u_x)_R  \right]  + \frac{\nu \alpha}{2 h}  ( u_R - u_L),
\label{upwnd_flux_diff}
\end{eqnarray}
$\alpha$ is a constant damping coefficient to be determined later, and $u_L$, $ (u_x)_L$, $u_R$, and $(u_x)_R$ are reconstructed point-valued solutions and derivatives at a face from the left and right cells, respectively. The left and right reconstructed solutions are evaluated by the 
$\kappa$-reconstruction scheme: Equations (\ref{kappa_uL_simple}) and (\ref{kappa_uR_simple}). As mentioned earlier, these solution values
are evaluated by reconstructed quadratic polynomials, e.g., $q_i(x)$ given in Equation (\ref{quadratic_poly_iw}) for $u_L$.  Then, it would be reasonable to compute $(u_x)_L$ by differentiating and evaluating $q_i(x)$: 
\begin{eqnarray}
(u_x)_L = \left.  \frac{d q_i(x)}{d x} \right|_{x=x_i+h/2} =  (u_x)_{i}   =  \frac{   \overline{u}_{i+1} -  \overline{u}_{i-1}  + 3 \kappa ( \overline{u}_{i+1} - 2 \overline{u}_{i}  +  \overline{u}_{i-1}  ) }{2 h},
\end{eqnarray}
which becomes for $\kappa=1/3$, 
\begin{eqnarray}
(u_x)_L   = \frac{  \overline{u}_{i+1} -  \overline{u}_{i}  }{h}.
\end{eqnarray}
Similarly, we find 
\begin{eqnarray}
(u_x)_R  = \left.   \frac{d q_{i+1}(x)}{d x} \right|_{x=x_{i+1}-h/2} =  (u_x)_{i+1} 
                 =  \frac{   \overline{u}_{i+2} -  \overline{u}_{i}  + 3 \kappa ( \overline{u}_{i+2} - 2 \overline{u}_{i+1}  +  \overline{u}_{i}  ) }{2 h}.
\end{eqnarray}
which again becomes for $\kappa=1/3$,
\begin{eqnarray}
 (u_x)_R = \frac{  \overline{u}_{i+1} -  \overline{u}_{i}  }{h}.
\end{eqnarray}
Therefore, the derivative of the quadratic reconstruction is continuous across the face. Then, the diffusion scheme is given by
\begin{eqnarray}
 - \frac{\nu}{h}  \left[  F^d_{i+1/2} -F^d_{i-1/2}  \right] = 
  - \nu \frac{  
 C_{i-2}  \overline{u}_{i-2} + C_{i-1}  \overline{u}_{i-1} + C_{i}  \overline{u}_{i}  + C_{i+1}  \overline{u}_{i+1}+ C_{i+2}  \overline{u}_{i+2} }{h^2},
\label{fv__diffusion_01}
\end{eqnarray}
where
\begin{eqnarray}
C_{i-2} = C_{i+2} = \frac{1}{8} \left[   \alpha (\kappa-1)  - 2 (  3 \kappa-1  ) \right], 
C_{i-1} = C_{i+1} = \frac{1}{8} \left[     24 \kappa  - 4 \alpha (\kappa-1) \right], 
C_i =   \frac{1}{8}  \left[   6  \alpha (\kappa-1) - 4 ( 9 \kappa + 1 )\right],
\label{diffusion_scheme_general_fomr}
\end{eqnarray}
which can be expanded as 
\begin{eqnarray}
 - \frac{\nu}{h}  \left[  F^d_{i+1/2} -F^d_{i-1/2}  \right] =  
- \nu u_{xx} -  \frac{1}{8}\left[  \alpha (\kappa-1)  - 3(2 \kappa-1)   \right] (\nu u_{xxxx}) h^2 + O(h^4),
\label{fv__diffusion_01_TE00}
\end{eqnarray}
thus leading to
\begin{eqnarray}
 - \frac{\nu}{h}  \left[  F^d_{i+1/2} -F^d_{i-1/2}  \right] =  
    \frac{1}{h} \int_{x_i - h/2}^{x_i+h/2} (  \nu  u_{xx}) \, dx
-   \frac{1}{8}\left[  \alpha (\kappa-1)  - 3(2 \kappa-1)   -   \frac{1}{3}  \right] (\nu u_{xxxx}) h^2 + O(h^4).
\label{fv__diffusion_01_TE}
\end{eqnarray}
Therefore, to achieve fourth-order accuracy, we must set
\begin{eqnarray}
   \alpha (\kappa-1)  - 3(2 \kappa-1)   -   \frac{1}{3}  = 0,
\end{eqnarray}
or
\begin{eqnarray}
   \alpha   =   \frac{  2  ( 4 - 9 \kappa ) }{ 3( 1- \kappa ) },
   \label{alpha_kappa_fourth}
\end{eqnarray}
giving for $\kappa=1/3$, 
\begin{eqnarray}
   \alpha   = 1
\end{eqnarray}
Note that it gives $\alpha = 8/3$ for $\kappa=0$, which is consistent with the value derived for fourth-order accuracy 
with $\kappa=0$ in Refs.\cite{nishikawa:AIAA2010,nishikawa_general_principle:CF2011}. 
Note also that the diffusion scheme (\ref{diffusion_scheme_wrong}) can be obtained with $\kappa=1/3$ and $\alpha=3/2$,
which does not satisfy the condition (\ref{alpha_kappa_fourth}).
It is interesting to note that the fourth-order diffusion scheme compatible with the MUSCL scheme 
is unique as the substitution of Equation (\ref{alpha_kappa_fourth}) into the scheme (\ref{diffusion_scheme_general_fomr}) gives
 \begin{eqnarray}
  - \frac{1}{h}  [ F^d_{i+1/2}  - F^d_{i-1/2} ]    =  - \nu \frac{  - u_{i-2} + 16 u_{i-1}  - 30 u_i + 16 u_{i+1} - u_{i+2}  }{12 h^2}.
  \label{diffusion_scheme_correct_third_order}
\end{eqnarray}
This is the diffusion scheme required to achieve third-order accuracy for the convection-diffusion equation (\ref{conv_diff_eq}).
 
\section{Numerical Results}
\label{results}


\subsection{Unsteady problem}

  \begin{figure}[t]
    \centering
      \hfill    
          \begin{subfigure}[t]{0.45\textwidth}
        \includegraphics[width=\textwidth]{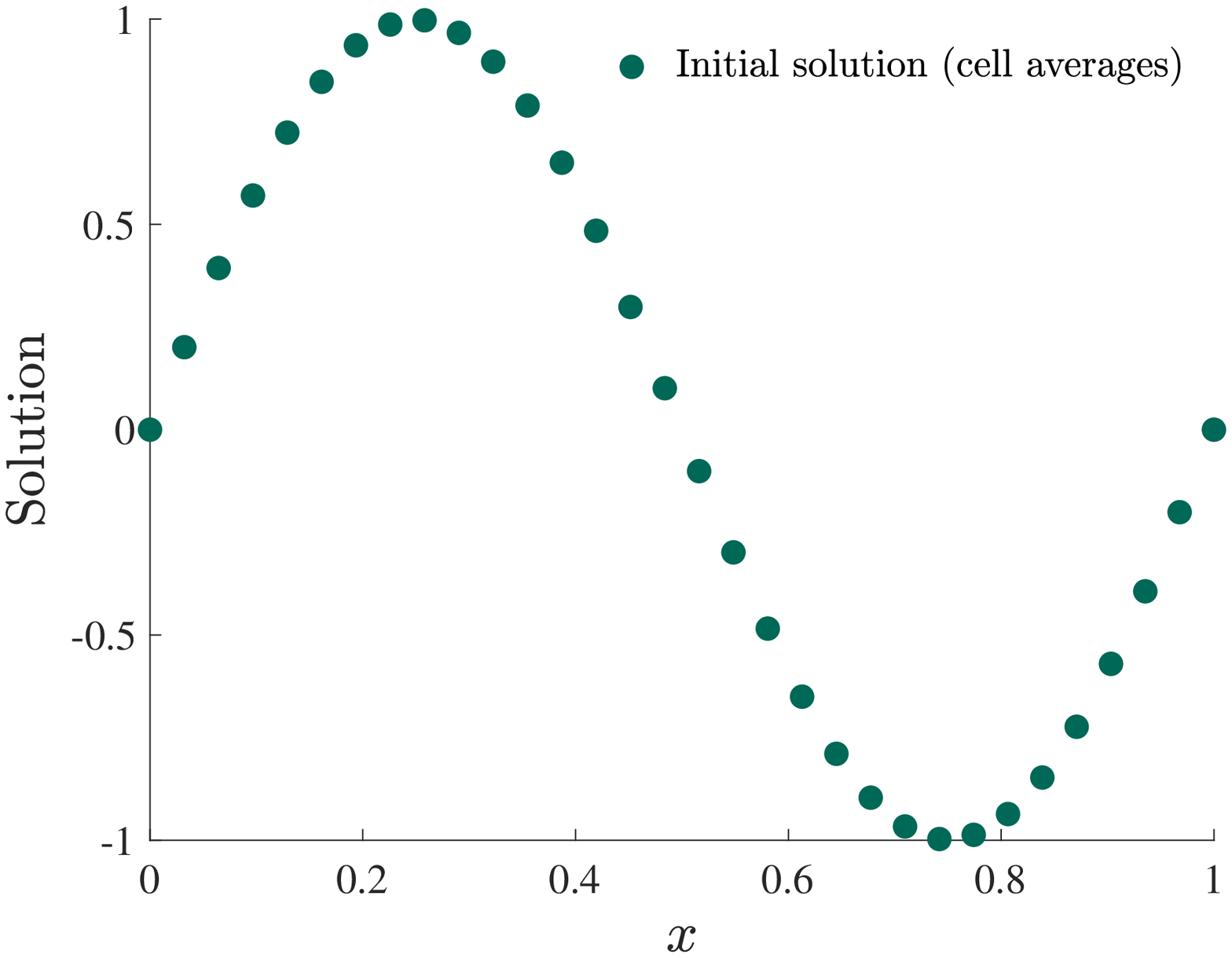}
          \caption{Initial solution (cell averages).}
          \label{fig:unsteady_initial}
      \end{subfigure}
      \hfill
          \begin{subfigure}[t]{0.45\textwidth}
        \includegraphics[width=\textwidth]{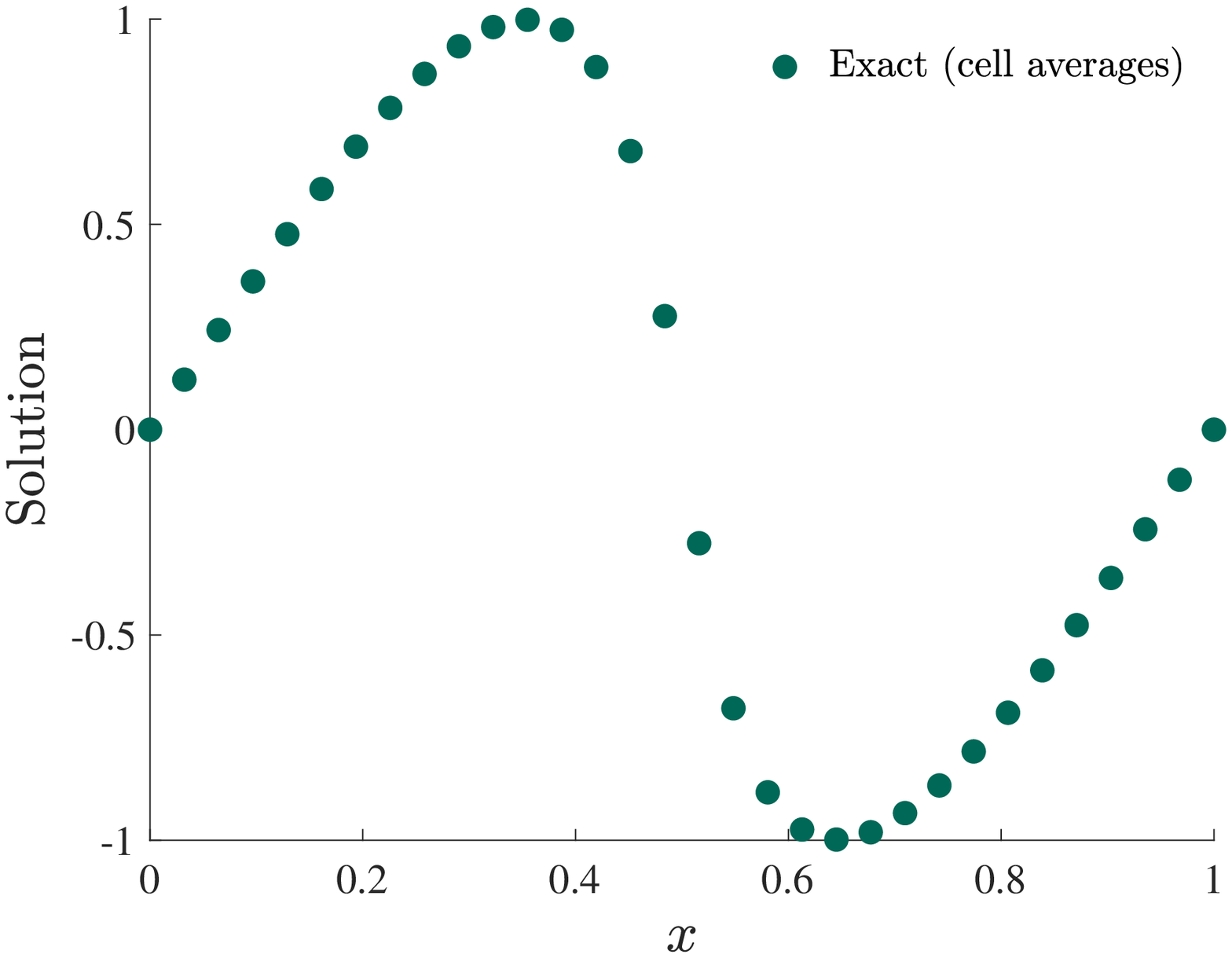}
          \caption{Final solution (cell averages).}
       \label{fig:unsteady_final}
      \end{subfigure}
      \hfill    
            \caption{
\label{fig:unsteady_solution}%
Initial and final solutions for the unsteady test case.
} 
\end{figure}

We consider a time-dependent problem for Burgers's equation: 
\begin{eqnarray}
u_t  + f_x= 0, 
\end{eqnarray}
where $f = u^2/2$ in $x \in [0,1]$ with the initial solution,
\begin{eqnarray}
u(x) = \sin (2 \pi x),
\end{eqnarray}
The domain is taken to be periodic (i.e., there is no boundary in this problem),
and the solution is computed at the final time $t = t_f =0.105$. The initial and final solutions are shown in Figure \ref{fig:unsteady_solution}. 
The dissipation coefficient $D$ is computed by the arithmetic average of the left and right values: $D = ( | u_L | + | u_R | ) 2$, which corresponds to 
the Roe linearization \cite{Roe_JCP_1981} with a smoothing technique of Harten and Hyman \cite{Harten_Hyman_JCP_1983} applied to $D$. 
The time integration is performed by
the three-stage SSP Runge-Kutta scheme \cite{SSP:SIAMReview2001} for the total of 840 time steps with 
a constant time step $\Delta t = 0.000125$, which has been found small enough for the discretization 
error (i.e., the error in the numerical solution) to be dominated by the spatial discretization.
To verify the spatial order of accuracy, we performed computations with $\kappa=0$, $1/2$, and $1/3$ for a series of grids: 32, 64, 128, 256, 512, 
1024, 2048 cells.

 Note that the initial solution must be cell-averaged since the numerical solution is 
a cell average in the MUSCL scheme. Thus, we set at the beginning, for each cell $i$, 
\begin{eqnarray}
\overline{u}_i  
=
\frac{1}{h} \int_{x_i - h/2}^{x_i+h/2} \sin (2 \pi x) \, dx  
=
\frac{1}{2 \pi h}  \left[  \cos \left(    \pi (  h-2 x_i)  \right)  -  \cos \left( \pi ( h+2 x_i)   \right)    \right]
=\frac{1}{ \pi h}  \sin( 2 \pi x_i ) \sin \left(     \pi h \right)
,
\end{eqnarray}
and then begin the time integration. The initial cell-averaged solution is plotted in Figure \ref{fig:unsteady_initial}. 
Note that if the initial solution is given by the point value $\overline{u}_i  =\sin (2 \pi x_i)$, then 
a second-order error will be introduced immediately, even before we begin the time integration. Later, we will demonstrate this numerically.

To emphasize the difference between the cell average and the point value, we measure the discretization error in three different ways, locally at a cell $i$, 
\begin{eqnarray}
{\cal E}_p (x_i) = \left|  \overline{u}_i - u_{exact}(x_i,t_f)   \right|, \quad 
{\cal E}_c (x_i) = \left|  \overline{u}_i - \frac{1}{h} \int_{x_i - h/2}^{x_i+h/2}  u_{exact}(x,t_f) \, dx    \right|, \quad
\hat{\cal E}_p (x_i) = \left|  \hat{u}_i - u_{exact}(x_i,t_f)   \right|,   
\end{eqnarray}
where $u_{exact}(x,t_f)$ is the exact point-valued solution at the final time,
\begin{eqnarray}
u_{exact}(x,t_f) =   \sin (2 \pi (x -  u_{exact} t) ),
\end{eqnarray}
which is defined implicitly and thus solved iteratively by the fix-point iteration (see Figure \ref{fig:unsteady_final} for the final solution plotted on the coarsest grid), and $ \hat{u}_i$ is a point-valued solution
at the cell center $x=x_i$ recovered from the cell-averaged solution $ \overline{u}_i$ as
\begin{eqnarray}
  \hat{u}_i = \overline{u}_i - \frac{1}{24} \left(  \frac{  \overline{u}_{i-1} -2  \overline{u}_i  +  \overline{u}_{i+1}   }{h^2} \right) h^2, 
  \label{point_recovered}
\end{eqnarray}
which is an accurate representation of the point value at $x=x_i$ up to fourth-order and thus sufficiently accurate to 
verify third-order accuracy of the numerical solution. Finally, we define the discretization error by the $L_1$ norm: 
\begin{eqnarray}
L_1( {\cal E}_p ) = \sum_{i=1}^n \frac{{\cal E}_p (x_i)  }{n} , \quad
L_1( {\cal E}_c ) = \sum_{i=1}^n \frac{{\cal E}_c (x_i)  }{n} , \quad
L_1( \hat{\cal E}_p ) = \sum_{i=1}^n \frac{ \hat{\cal E}_p (x_i)  }{n} , 
\end{eqnarray}
where $n$ is the number of cells in a grid.

  \begin{figure}[th!]
    \centering
      \hfill    
                \begin{subfigure}[t]{0.32\textwidth}
        \includegraphics[width=\textwidth]{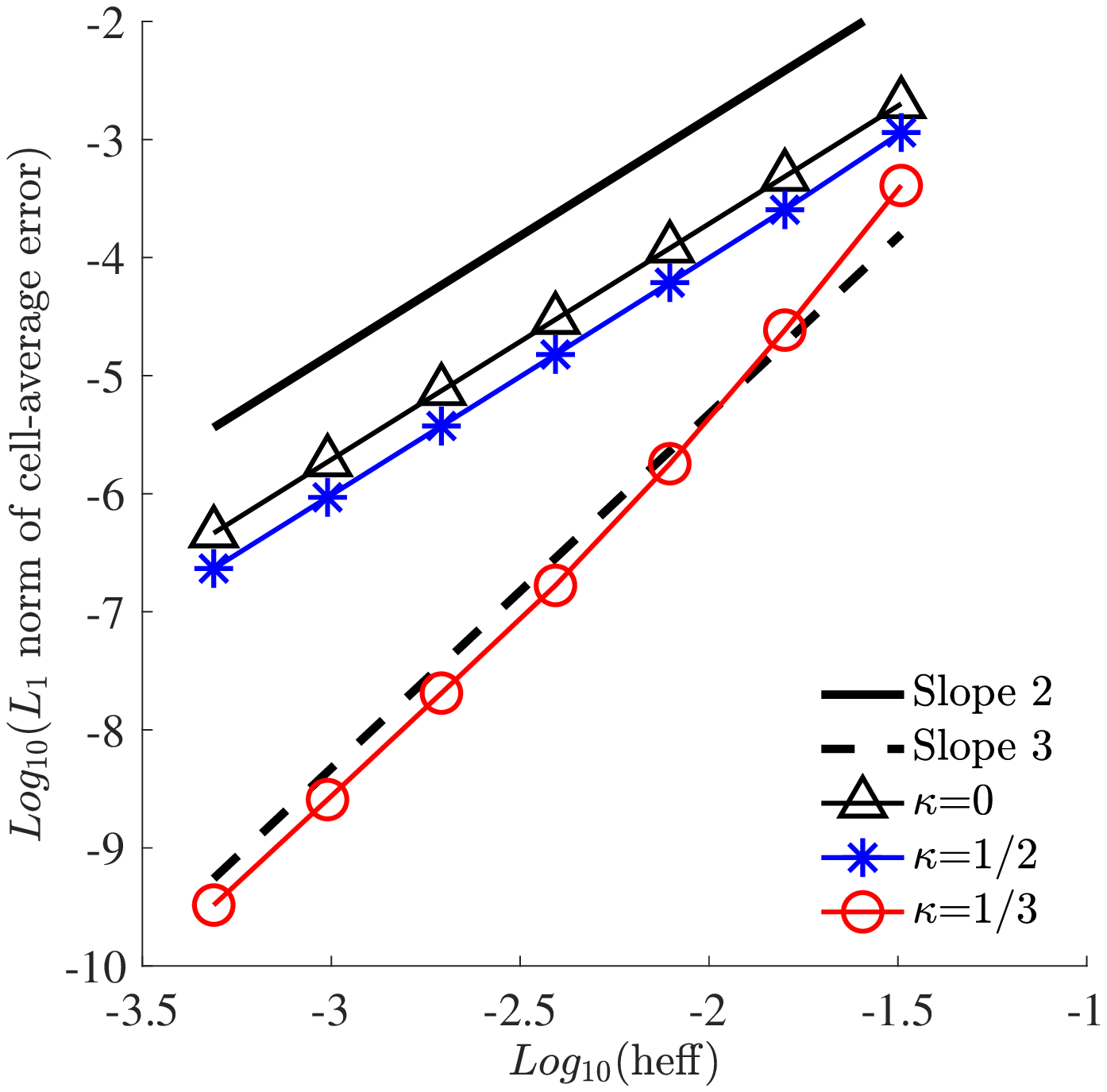}
          \caption{Error convergence for $L_1( {\cal E}_c) $.}
       \label{fig:unsteady_err_ca}
      \end{subfigure}
      \hfill
          \begin{subfigure}[t]{0.32\textwidth}
        \includegraphics[width=\textwidth]{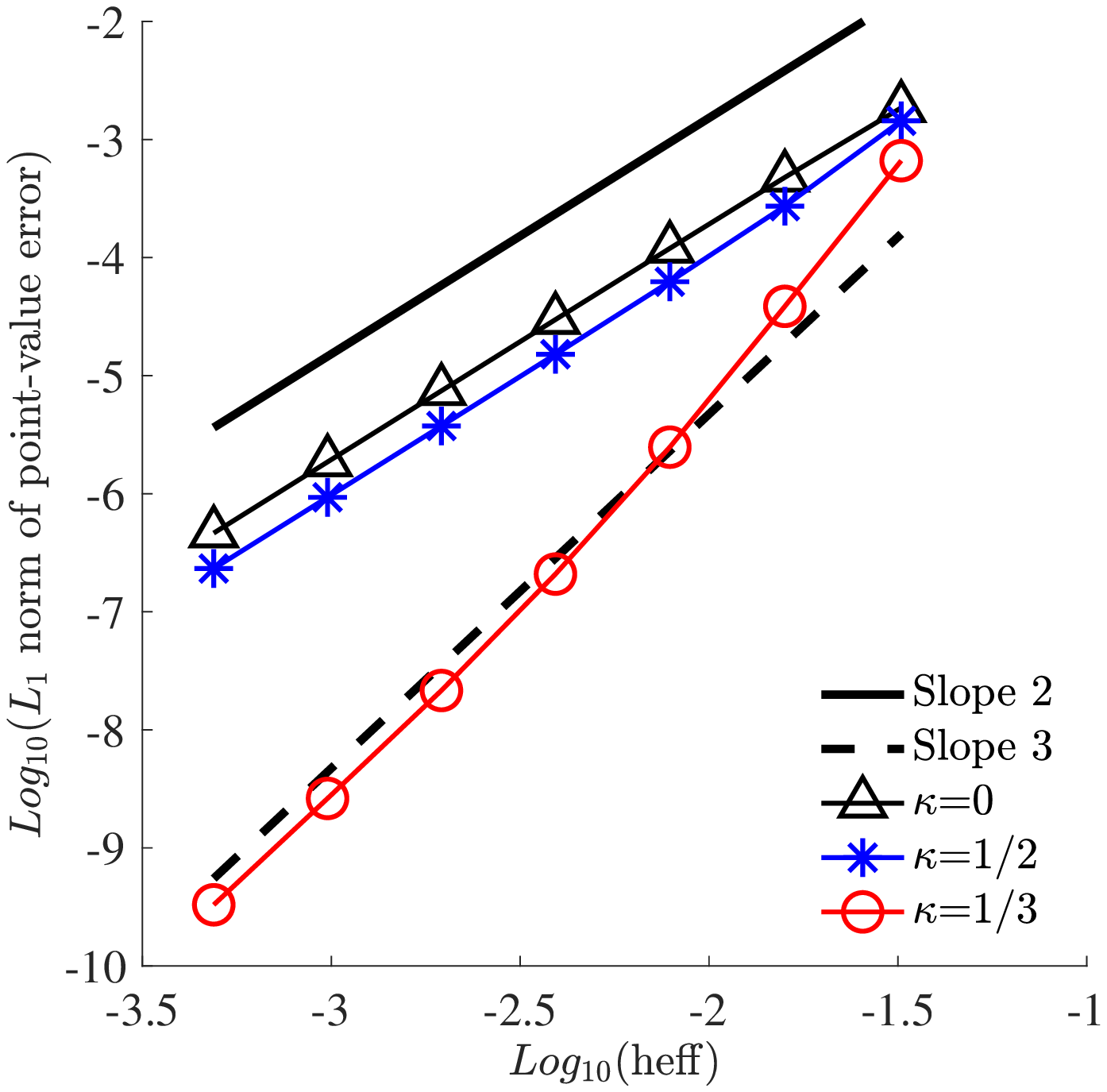}
          \caption{Error convergence for $L_1( \hat{\cal E}_p ) $.}
       \label{fig:unsteady_err_pca}
      \end{subfigure}
      \hfill    
          \begin{subfigure}[t]{0.32\textwidth}
        \includegraphics[width=\textwidth]{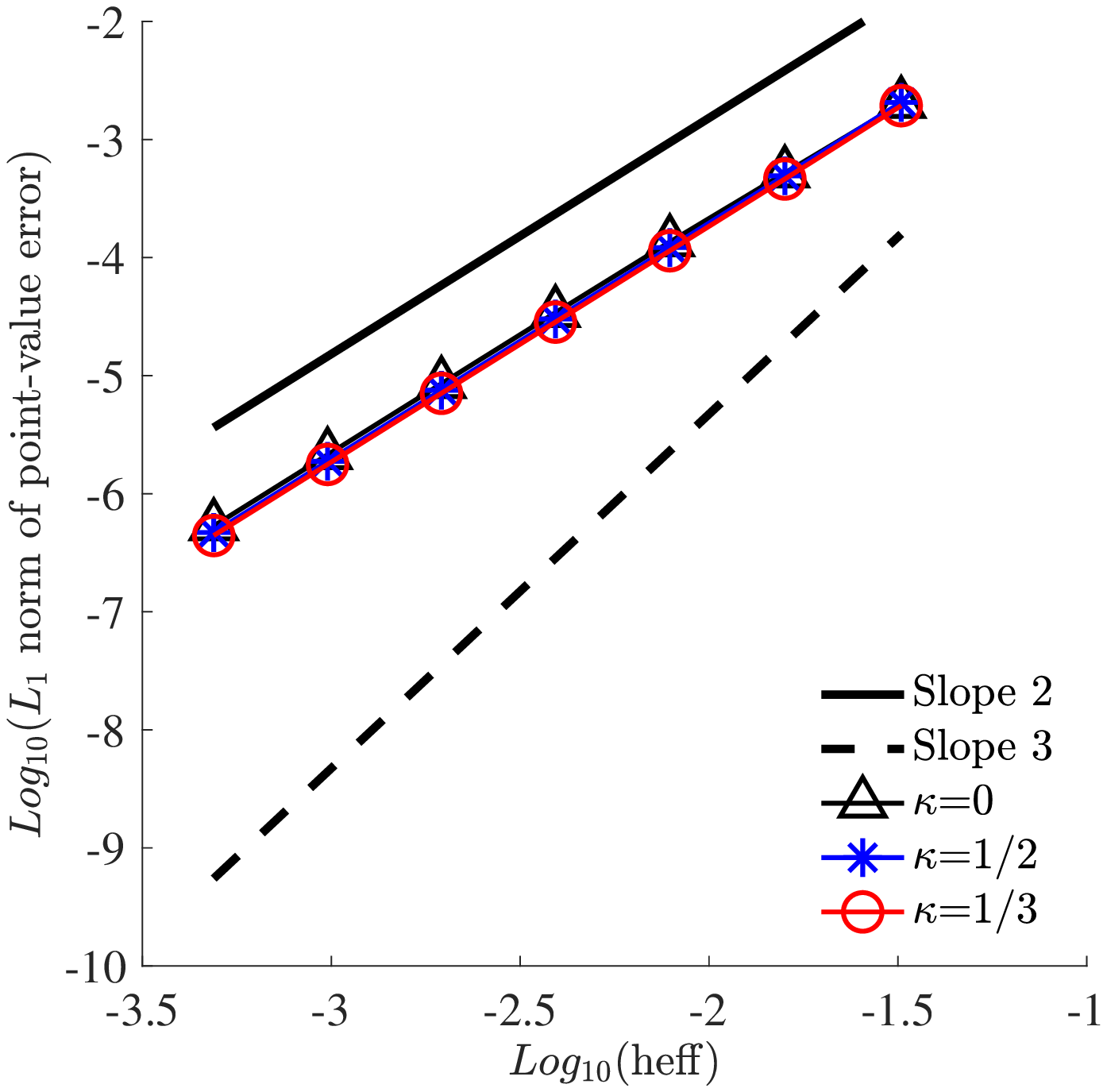}
          \caption{Error convergence for $L_1( {\cal E}_p ) $.}
          \label{fig:unsteady_err_p}
      \end{subfigure}
      \hfill
            \caption{
\label{fig:unsteady_error}%
Error convergence results for the unsteady test case: cell-averaged initial solutions.
} 
\end{figure}
  \begin{figure}[th!]
    \centering
      \hfill    
                \begin{subfigure}[t]{0.32\textwidth}
        \includegraphics[width=\textwidth]{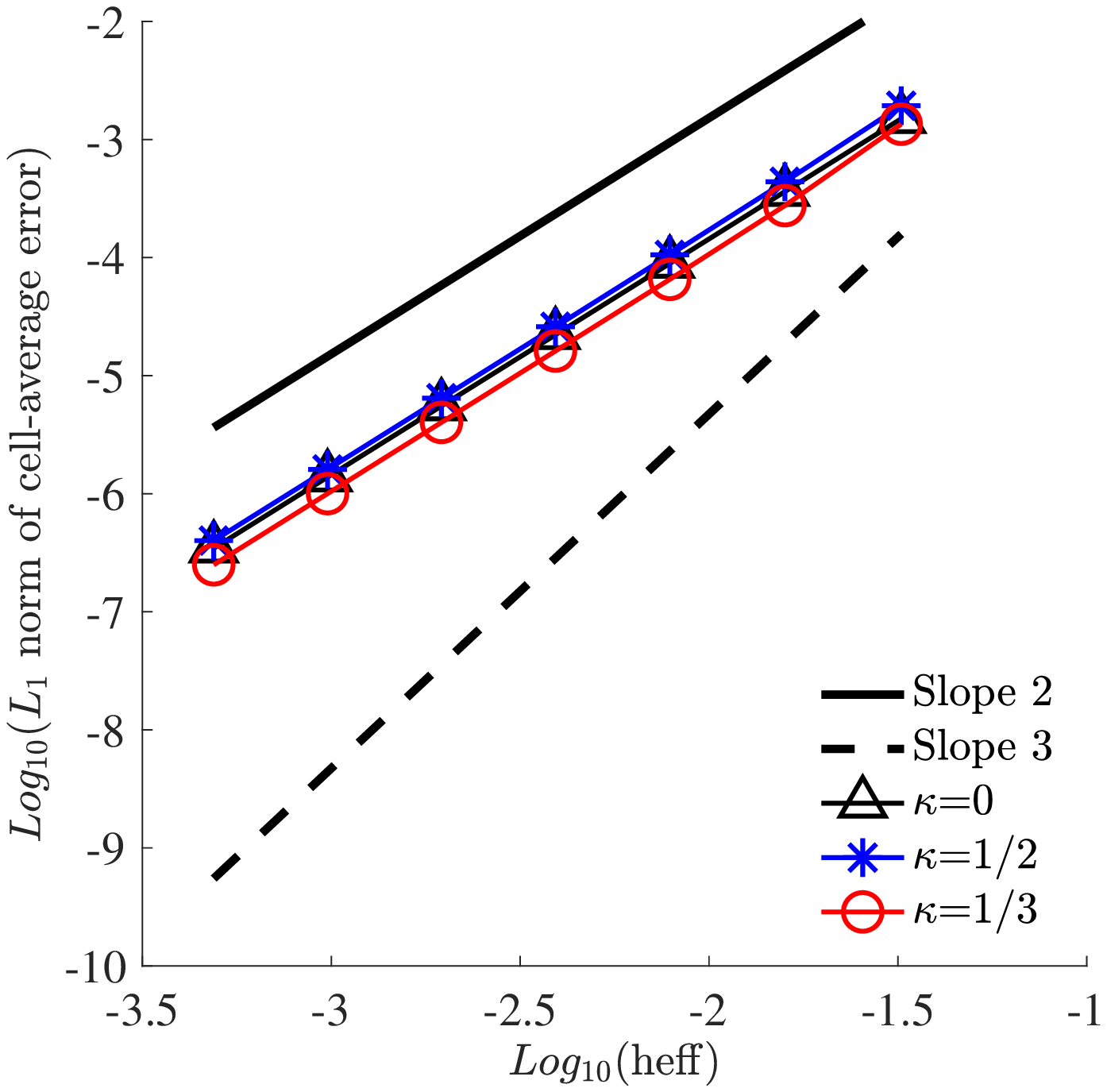}
          \caption{Error convergence for $L_1( {\cal E}_c) $.}
       \label{fig:unsteady_err_ca_pi}
      \end{subfigure}
      \hfill
          \begin{subfigure}[t]{0.32\textwidth}
        \includegraphics[width=\textwidth]{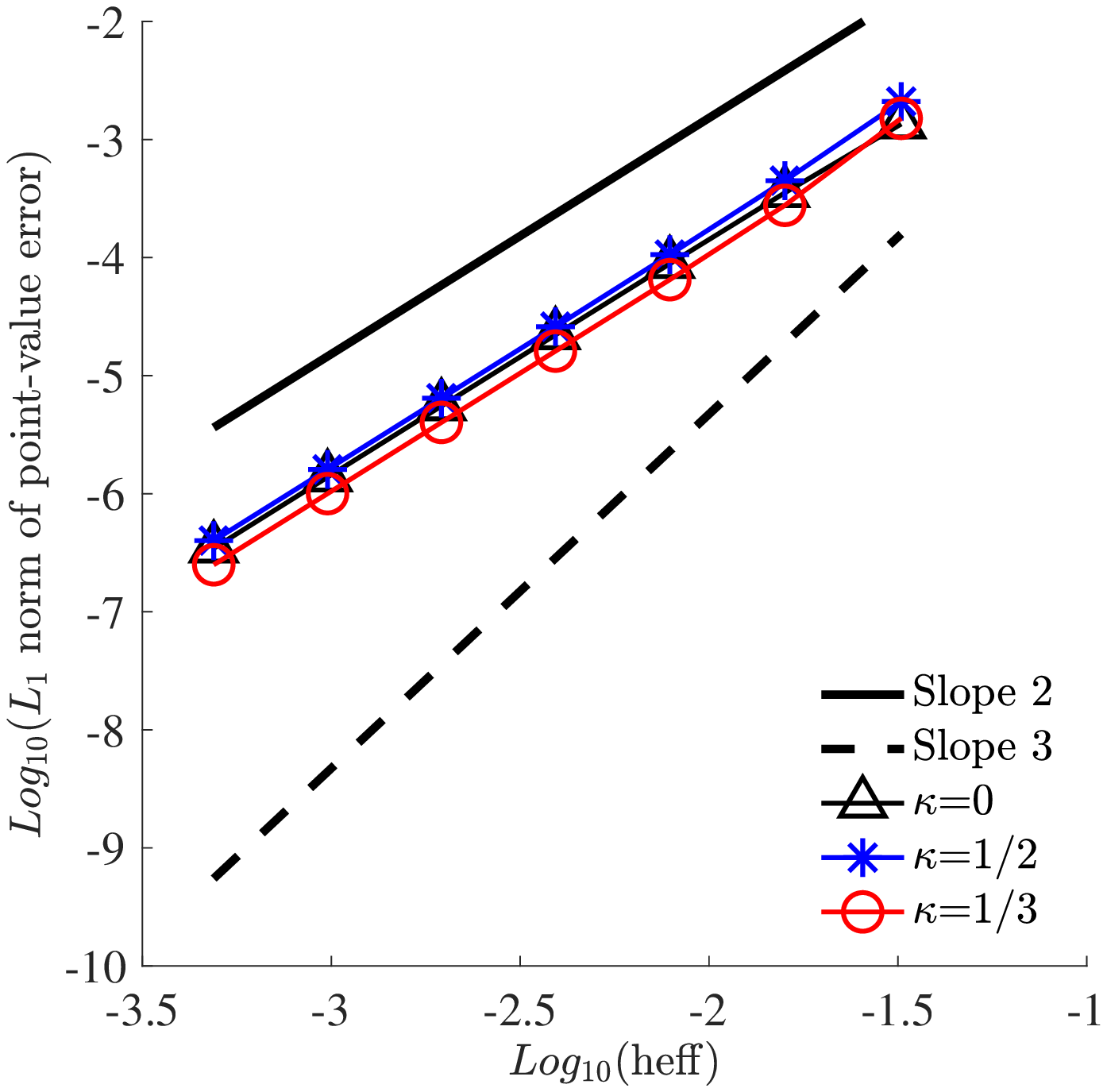}
          \caption{Error convergence for $L_1( \hat{\cal E}_p ) $.}
       \label{fig:unsteady_err_pca_pi}
      \end{subfigure}
      \hfill    
          \begin{subfigure}[t]{0.32\textwidth}
        \includegraphics[width=\textwidth]{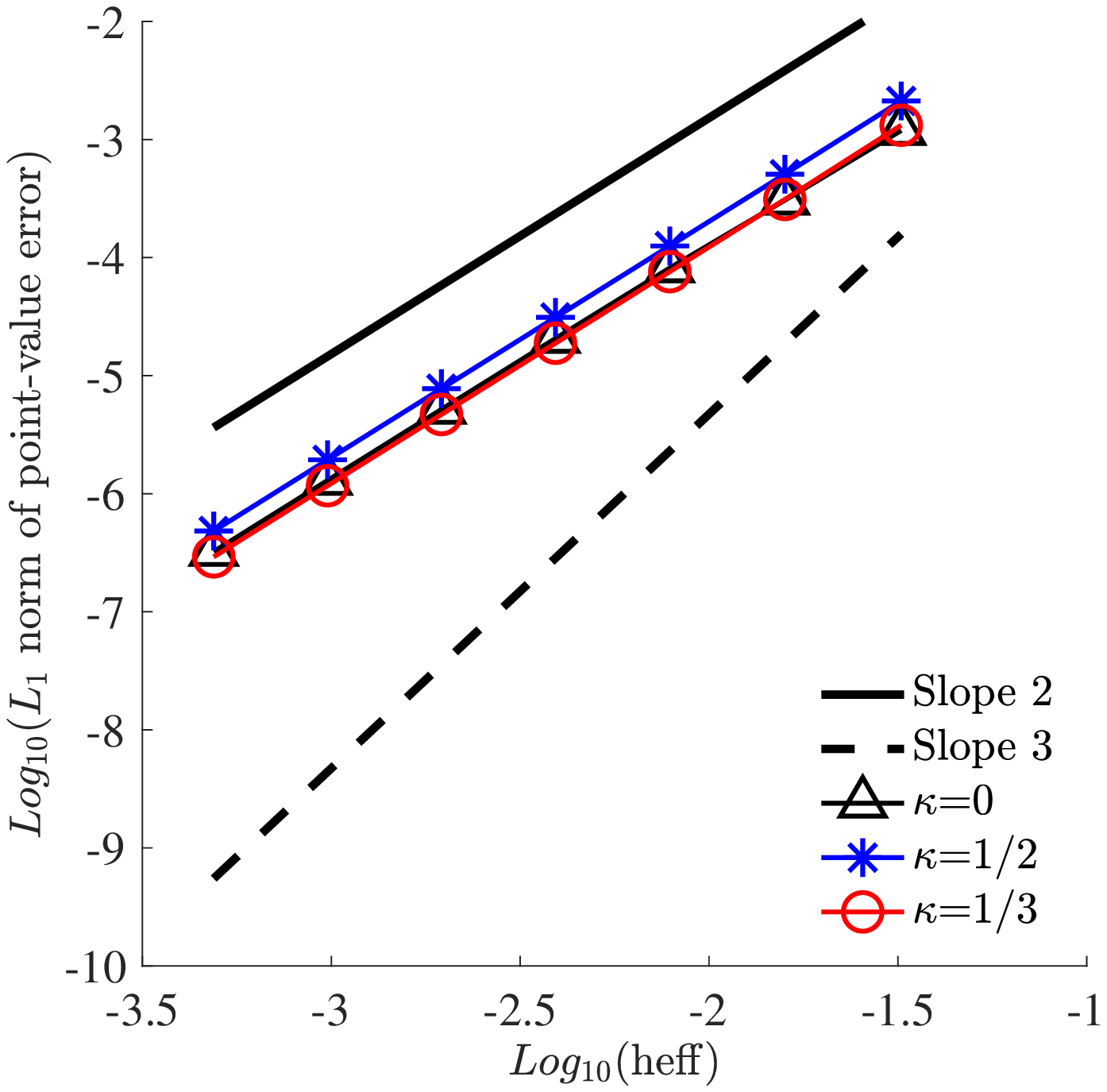}
          \caption{Error convergence for $L_1( {\cal E}_p ) $.}
          \label{fig:unsteady_err_p_pi}
      \end{subfigure}
      \hfill
            \caption{
\label{fig:unsteady_error_p}%
Error convergence results for the unsteady test case: point-valued initial solution
} 
\end{figure}
  \begin{figure}[th!]
    \centering
      \hfill    
                \begin{subfigure}[t]{0.32\textwidth}
        \includegraphics[width=\textwidth]{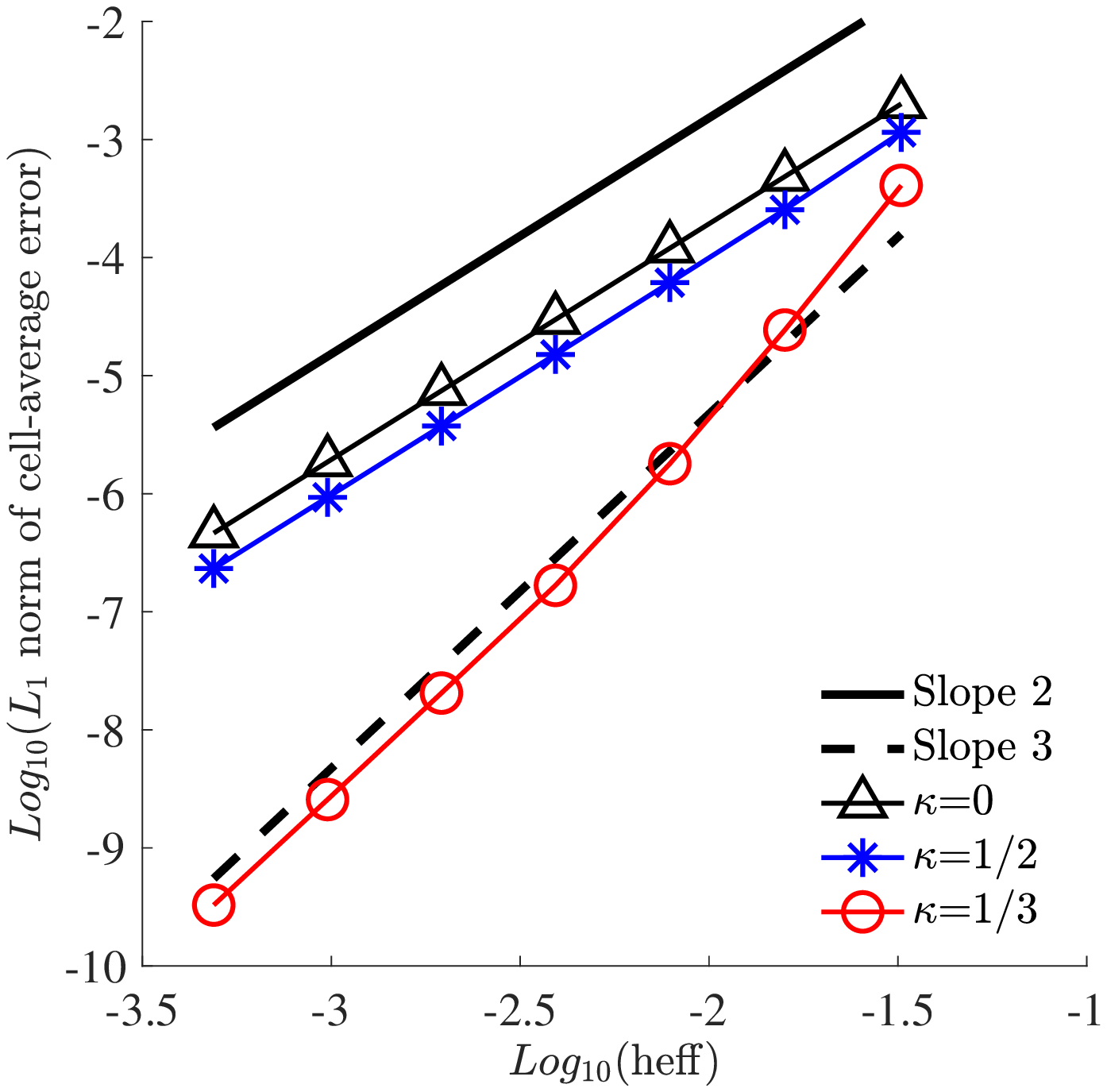}
          \caption{Error convergence for $L_1( {\cal E}_c) $.}
       \label{fig:unsteady_err_ca_pic}
      \end{subfigure}
      \hfill
          \begin{subfigure}[t]{0.32\textwidth}
        \includegraphics[width=\textwidth]{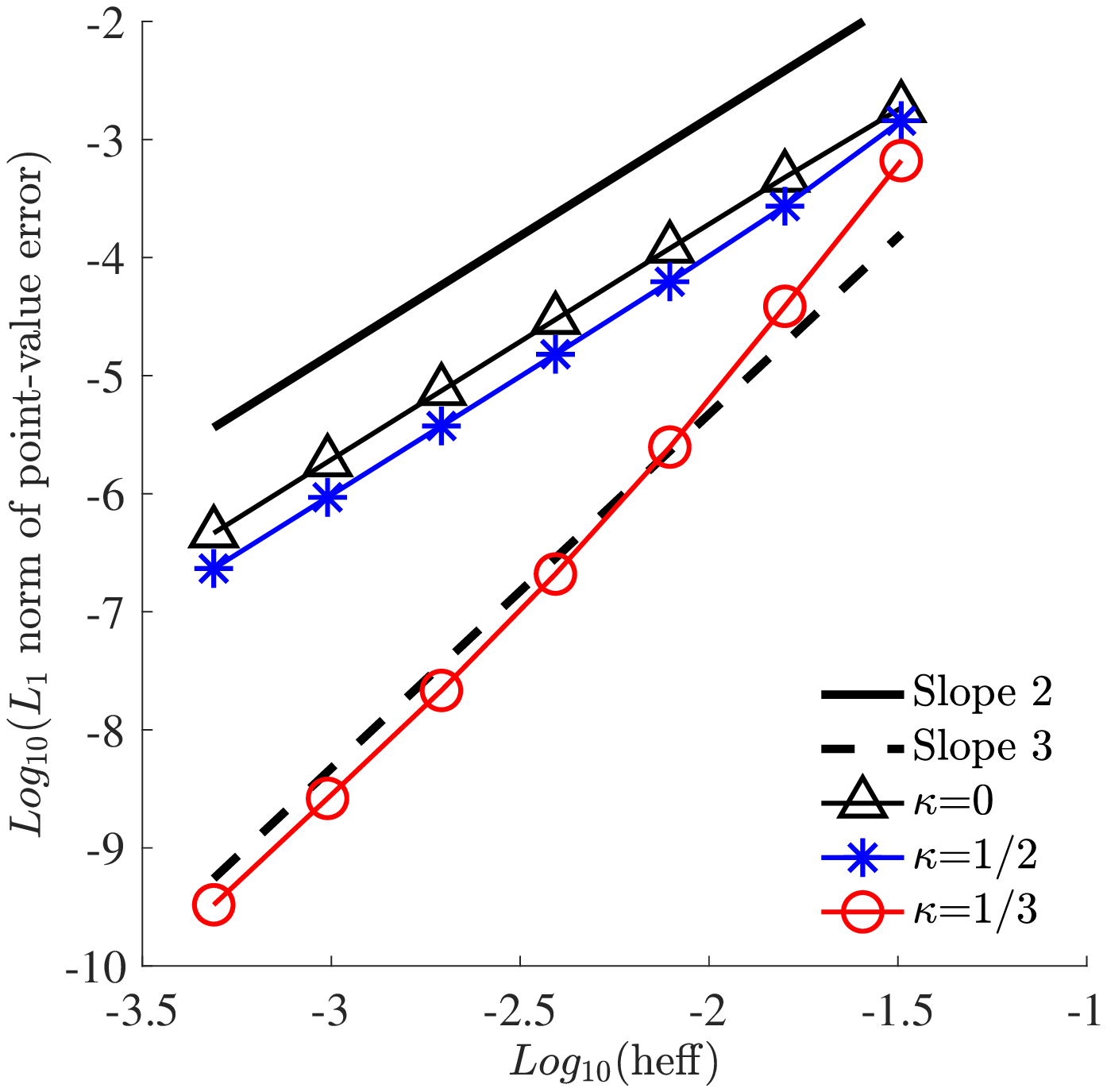}
          \caption{Error convergence for $L_1( \hat{\cal E}_p ) $.}
       \label{fig:unsteady_err_pca_pic}
      \end{subfigure}
      \hfill    
          \begin{subfigure}[t]{0.32\textwidth}
        \includegraphics[width=\textwidth]{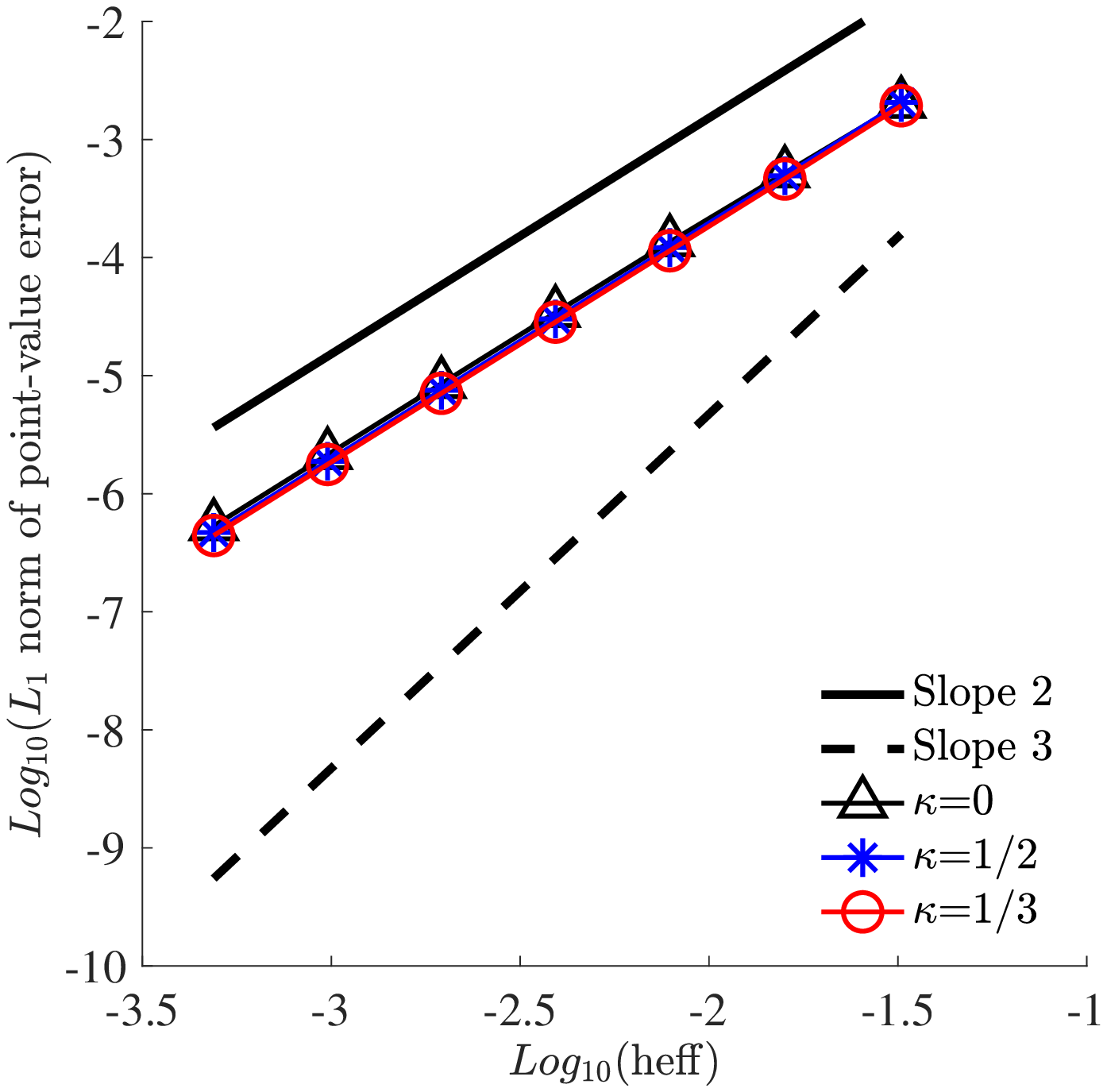}
          \caption{Error convergence for $L_1( {\cal E}_p ) $.}
          \label{fig:unsteady_error_c}
      \end{subfigure}
      \hfill
            \caption{
\label{fig:unsteady_error_point_c}%
Error convergence results for the unsteady test case: corrected point-valued initial solution.
} 
\end{figure}
Error convergence results are shown in Figure \ref{fig:unsteady_error} for the case of cell-averaged initial solutions.
As can be seen clearly in Figure \ref{fig:unsteady_err_ca}, the numerical solution is third-order accurate as a cell-averaged solution with $\kappa=1/3$ as expected.
The choice $\kappa=1/2$, which corresponds to the QUICK interpolation scheme, leads to second-order accuracy
although slightly more accurate than Fromm's scheme ($\kappa=0$). Figure \ref{fig:unsteady_err_pca} shows the result for the point-valued solution recovered from 
the cell-averaged solution as in Equation (\ref{point_recovered}). Clearly, the recovered point-value solution is third-order 
accurate for $\kappa=1/3$, but remains second-order accurate for the other choices. 
 Finally, Figure \ref{fig:unsteady_err_p} shows the error convergence for the pointwise error norm $L_1( {\cal E}_p )$. As expected, the numerical solution is second-order accurate as a point value for any $\kappa$.
 
 To illustrate the importance of using cell-averaged initial solutions, 
we performed the same numerical experiments with point-valued initial solutions: $\overline{u}_i  =\sin (2 \pi x_i)$ at $t=0$ for all cells, $i=1,2,3, \cdots, n$. 
 Error convergence results are shown in Figure \ref{fig:unsteady_error_p}. As expected, the error is second-order for all error norms, indicating that the error is dominated by the initial second-order error committed in the initial solution. If we wish to avoid computing the cell-averaged initial solution, we may compute a cell-averaged initial solution from a point-valued initial solution with sufficient accuracy by
 \begin{eqnarray}
  \overline{u}_i = {u}_i  + \frac{1}{24} \left(  \frac{  {u}_{i-1} -2  {u}_i  +  {u}_{i+1}   }{h^2} \right) h^2.
  \label{cell_average_recovered}
\end{eqnarray}
In this way, the initial solution does not been to be integrated over each cell.
Results obtained with this correction are shown in Figure \ref{fig:unsteady_error_point_c}. As can be seen in 
Figures \ref{fig:unsteady_err_ca_pic} and \ref{fig:unsteady_err_pca_pic}, third-order accuracy is achieved 
for the cell-averaged solution and the recovered point-value solution.

\subsection{Steady convection problem}
\label{results_steady_conv}
 
To demonstrate third-order accuracy for a convection equation with a forcing term, we 
 consider a steady problem for Burgers's equation in $x \in [0,1]$: 
\begin{eqnarray}
f_x =  s(x),
\end{eqnarray}
where $f = u^2/2$, with the forcing term, 
\begin{eqnarray}
s(x) =  2  \sin (2  x) \cos(  2 x ),
\end{eqnarray}
so that the exact solution is given by
\begin{eqnarray}
u(x) = \sin (2  x).
\end{eqnarray}
For this problem, we drop the time-derivative term and define the residual at a cell $i$ as
\begin{eqnarray}
Res_i =   \frac{1}{h} [ F^c_{i+1/2}  - F^c_{i-1/2} ]   - \overline{s}_i,
\label{steady_residual}
\end{eqnarray}
where the forcing term is exactly cell-averaged over each cell:
\begin{eqnarray}
  \overline{s}_i   =   \frac{1}{h} \int_{x_i-h/2}^{x_i+h/2} {s}(x) \, dx 
  =  \frac{1}{2h} \left[ 
    \cos^2(  h - 2 x_i  ) -    \cos^2(  h + 2 x_i  ) 
  \right].
\end{eqnarray}
Then, we solve the system of nonlinear residual equations for the numerical solution: 
$\overline{u}_3, \overline{u}_4,  \cdots, \overline{u}_{n-3}, \overline{u}_{n-2}$.
Note that we will provide the exact cell-averaged solutions at the left two cells $i=1$ and $2$, and at the 
right cells $i=n-1$ and $n$, in order to exclude boundary effects, which are beyond the scope of the present study.
An implicit solver based on the exact Jacobian of the first-order scheme is used to solve the residual equations.
See Ref.\cite{nishikawa_liu_jcp2018}, for example, for further details of the implicit solver for a one-dimensional finite-volume scheme. 
To verify the order of accuracy, we solve the steady problem over a series of grids with 15, 31, 63, 127 cells.
As in the unsteady case, we consider two discretization error norms:
\begin{eqnarray}
L_1( {\cal E}_p) = \frac{1}{n-4} \sum_{i=3}^{n-2} |  \overline{u}_i - u_i^{exact} |,  \quad
L_1( {\cal E}_c) = \frac{1}{n-4} \sum_{i=3}^{n-2} | \overline{u}_i - \overline{u}_i^{exact}|,
\end{eqnarray}
where $ \overline{u}_i^{exact}$ is the exact cell-averaged solution:
\begin{eqnarray}
\overline{u}_i^{exact}
=
\frac{1}{h} \int_{x_i - h/2}^{x_i+h/2} \sin (2 x) \, dx  
=
\frac{1}{2  h}  \left[  \cos \left(      h-2 x_i   \right)  -  \cos \left(  h+2 x_i   \right)    \right]
=\frac{1}{ 2h}  \sin( 2  x_i ) \sin \left(   h \right).
\end{eqnarray}

We also verify the order of truncation error numerically by substituting the exact solution into the residual
(\ref{steady_residual}), and taking the $L_1$ norm over the cells. We consider both the point-valued and cell-averaged 
solutions, and define the following two truncation error norms:
\begin{eqnarray}
L_1( {\cal T}_p) = \frac{1}{n-4} \sum_{i=3}^{n-2} Res_i( \{ u_i^{exact} \} ), \quad
L_1( {\cal T}_c) = \frac{1}{n-4} \sum_{i=3}^{n-2} Res_i( \{ \overline{u}_i^{exact} \} ),
\end{eqnarray}
where $Res_i( \{ u_i^{exact} \} )$ is the residual with the point-valued exact solution substituted, 
and $Res_i( \{ \overline{u}_i^{exact} \} )$ is the residual with the cell-averaged exact solution substituted,

  \begin{figure}[th!]
    \centering
      \hfill  
                \begin{subfigure}[t]{0.32\textwidth}
        \includegraphics[width=\textwidth]{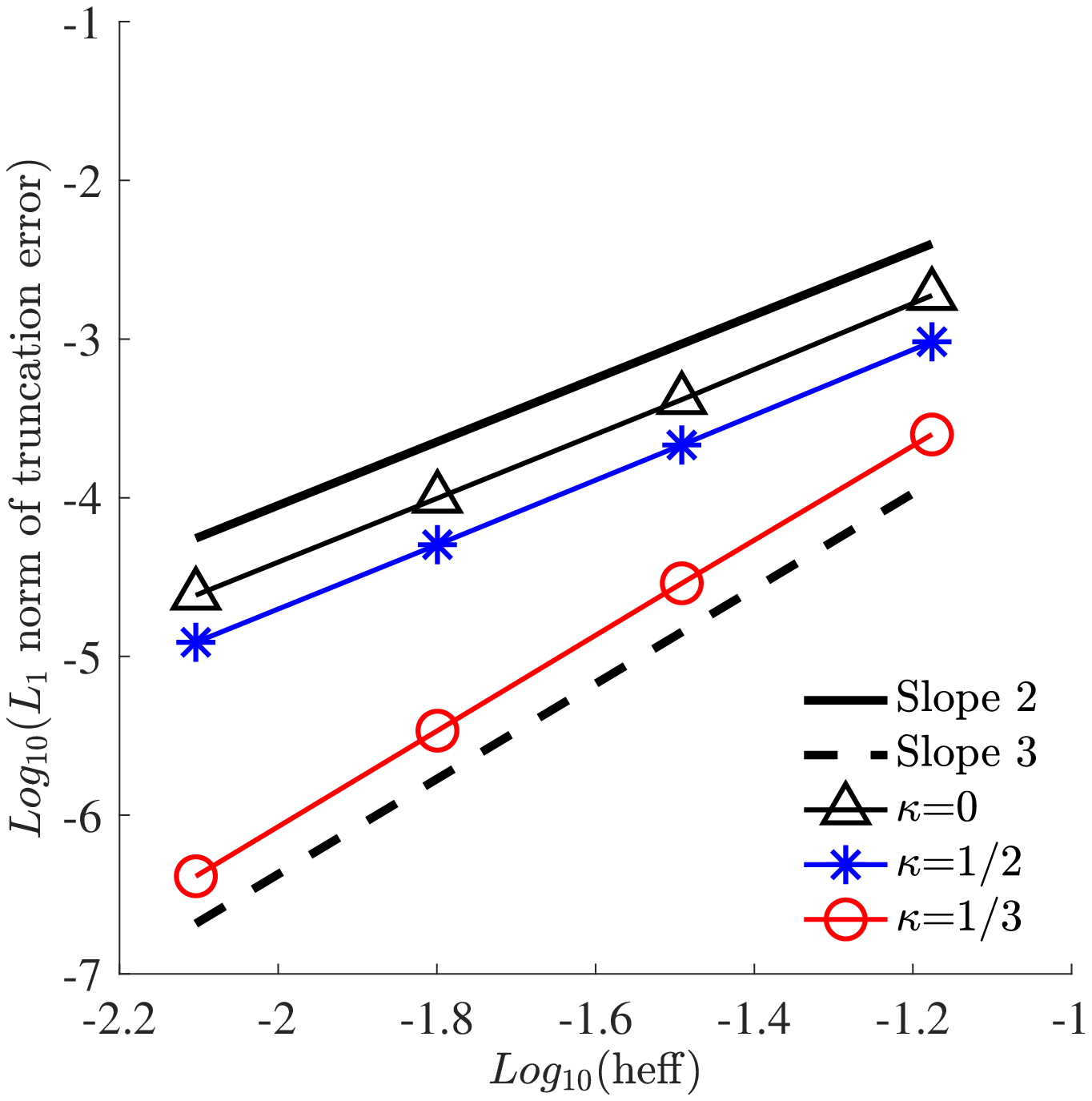}
          \caption{ $L_1( {\cal T}_c) $.}
       \label{fig:steady_adv_te_c}
      \end{subfigure}
      \hfill
          \begin{subfigure}[t]{0.32\textwidth}
        \includegraphics[width=\textwidth]{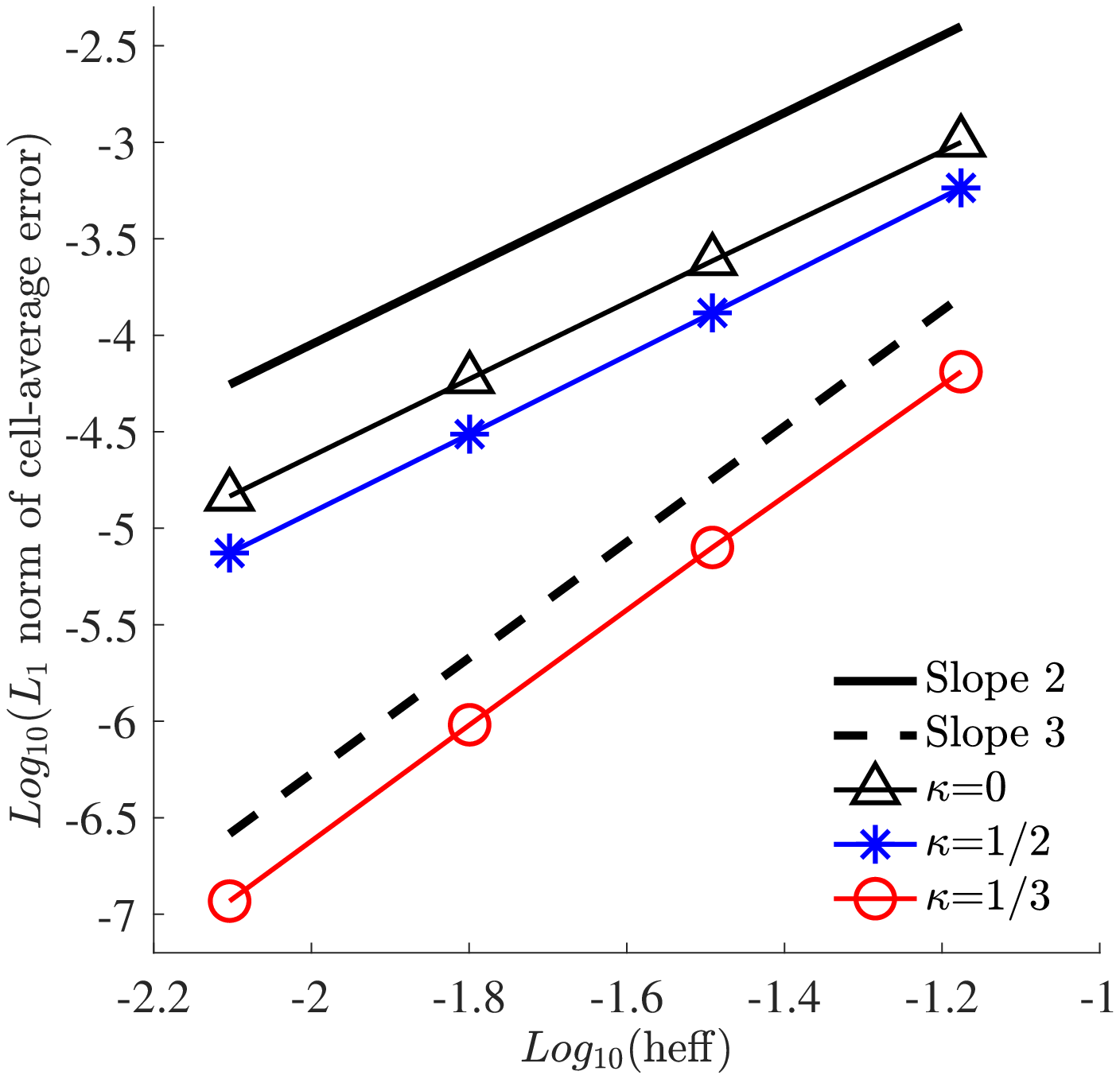}
          \caption{ $L_1( {\cal E}_c ) $.}
       \label{fig:steady_adv_err_ca}
      \end{subfigure}
      \hfill
          \begin{subfigure}[t]{0.32\textwidth}
        \includegraphics[width=\textwidth]{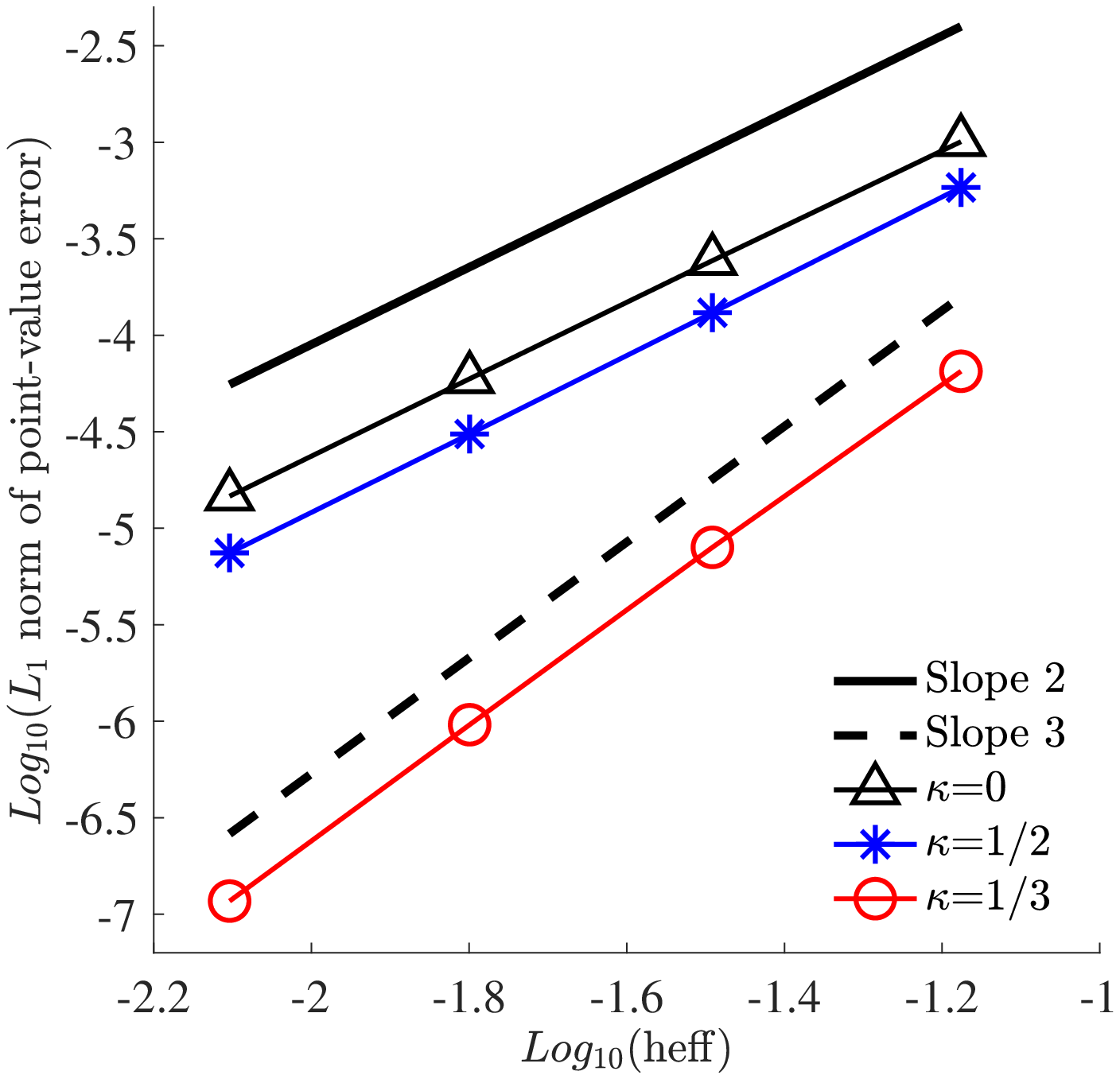}
          \caption{ $L_1( \hat{\cal E}_p ) $.}
       \label{fig:steady_adv_err_ca}
      \end{subfigure}
      \hfill  
      \\
      \hfill    
      \begin{center}
                \begin{subfigure}[t]{0.32\textwidth}
        \includegraphics[width=\textwidth]{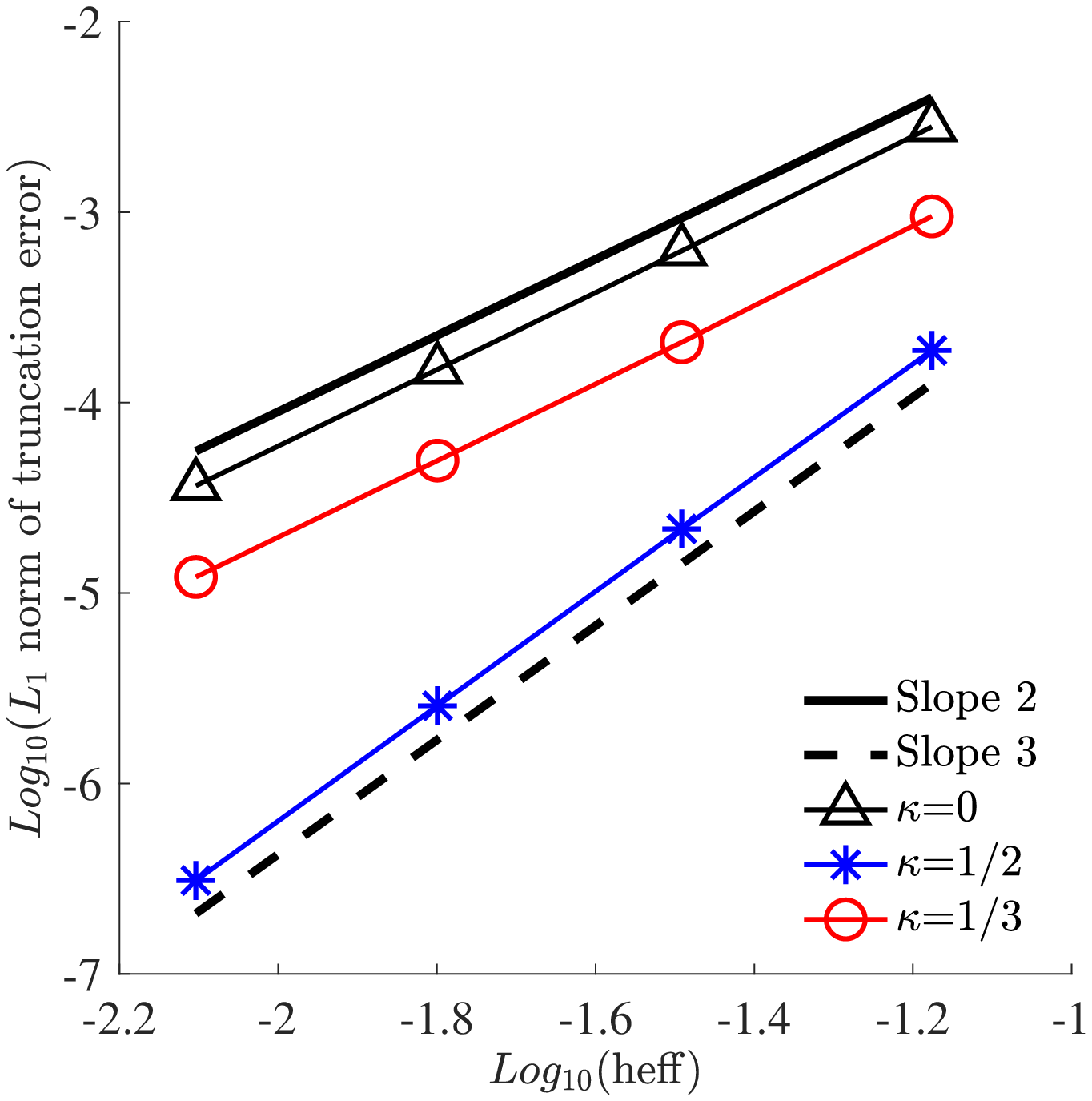}
          \caption{ $L_1( {\cal T}_p) $.}
       \label{fig:steady_adv_te_p}
      \end{subfigure}
                \begin{subfigure}[t]{0.32\textwidth}
        \includegraphics[width=\textwidth]{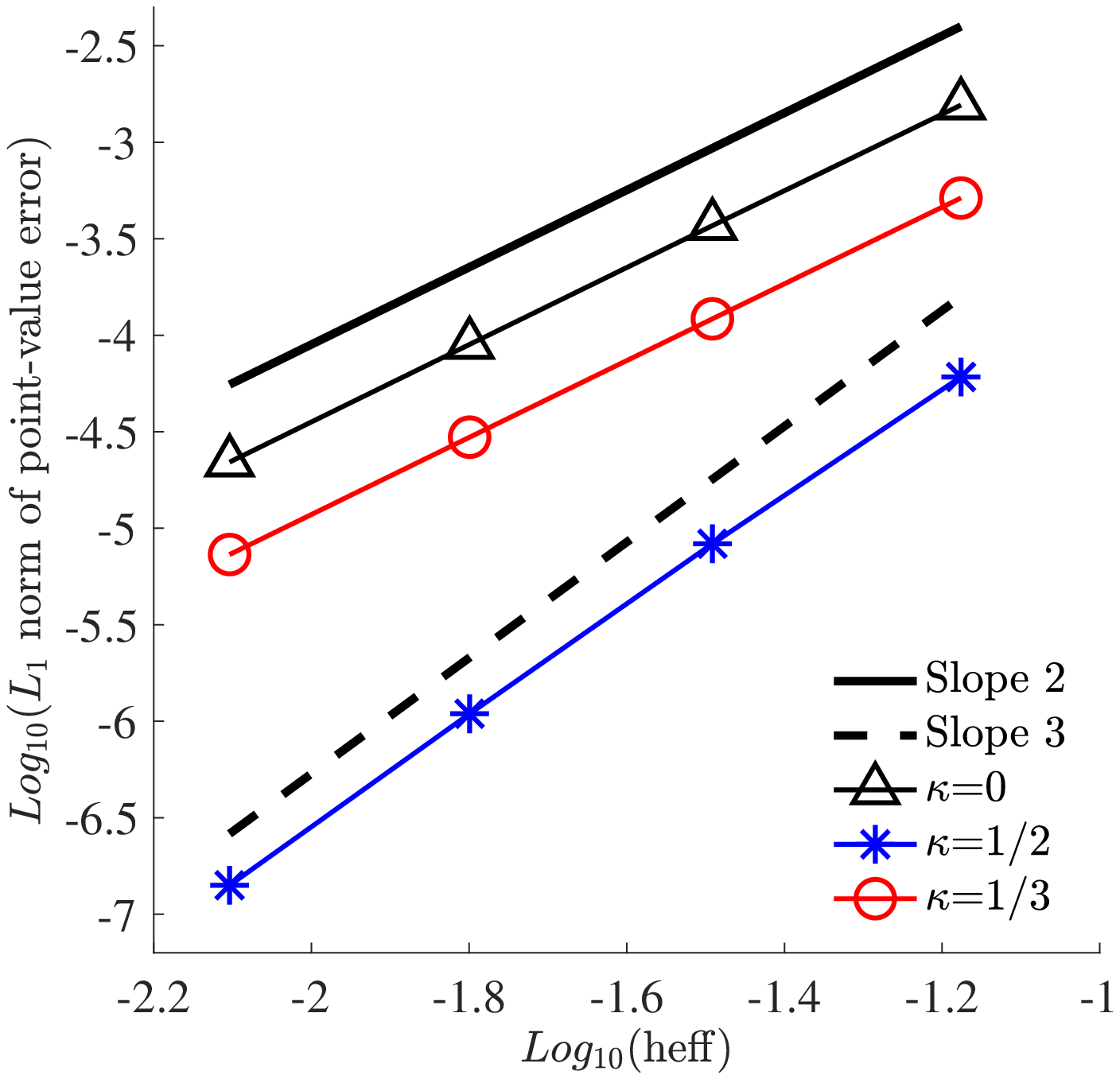}
          \caption{ $L_1( {\cal E}_p) $.}
       \label{fig:steady_adv_err_p}
      \end{subfigure}
      \end{center}
      \hfill    
      \\
            \caption{
\label{fig:steady_adv_error}%
Truncation and discretization error convergence results for the case of the steady Burgers equation.
} 
\end{figure}

Error convergence results are shown in Figure \ref{fig:steady_adv_error}. We first consider our target case, where the errors are
measured with the exact cell-averaged solution. Figure \ref{fig:steady_adv_te_c} shows the truncation error convergence for 
$L_1( {\cal T}_c)$. As expected, it is third-order only with $\kappa=1/3$.
The discretization error is also third-order in both the cell-averaged solution and the recovered point-value solution.
These results confirm our truncation error analysis. 

On the other hand, if we compute the truncation error with the point-valued exact solution, we see third-order accuracy achieved
with $\kappa=1/2$ as shown in Figure \ref{fig:steady_adv_te_p}. The same is observed for the discretization error. 
See Figure \ref{fig:steady_adv_err_p}. To understand the result, we will need to clarify the QUICK scheme ($\kappa=1/2$), which is planned 
in a subsequent paper. Therefore, here, we give only a short explanation. First of all, the numerical solution for a steady problem can be interpreted as
either a point value or a cell average since it does not depend on an initial solution. Then, the flux balance term in the steady residual (\ref{steady_residual}) is third-order with 
either combination: the $\kappa=1/3$ quadratic reconstruction with cell-averaged solutions or the $\kappa=1/2$ quadratic interpolation with point-valued solutions. The latter is the QUICK scheme of Leonard \cite{Leonard_QUICK_CMAME1979}, as mentioned earlier; it is third-order for this 
steady convection equation with the cell-averaged forcing term. A further discussion on the QUICK scheme will be given in a subsequent paper.
It is pointed out that third-order with $\kappa=1/2$ is a special case and does not hold for a convection-diffusion problem, as we will
discuss in the next section.

\subsection{Steady convection-diffusion problem}
\label{results_steady_conv_diff}
 
To demonstrate the importance of the diffusion scheme, we consider a steady problem for the viscous Burgers equation with a forcing term: 
\begin{eqnarray}
f_x =  \nu u_{xx} +s(x),
\end{eqnarray}
where $f = u^2/2$ and the forcing term is given by
\begin{eqnarray}
s(x) =  2  \sin (2 x) \cos(  2 x ) + 4 \nu \sin( 2 x),
\end{eqnarray}
so that the exact solution is given by
\begin{eqnarray}
u(x) = \sin (2  x).
\end{eqnarray}
Again, we integrate the forcing term is integrated exactly over each cell:
\begin{eqnarray}
  \overline{s}_i   =   \frac{1}{h} \int_{x_i-h/2}^{x_i+h/2} {s}(x) \, dx 
  =  \frac{1}{2h} \left[ 
    \cos^2(  h - 2 x_i  ) -    \cos^2(  h + 2 x_i  ) 
  \right] 
-  \frac{2 \nu }{h} \left[ 
    \cos(  h - 2 x_i  ) -    \cos(  h + 2 x_i  ) 
  \right].
\end{eqnarray}
In particular, we consider the case $\nu=1$, for which the convective and diffusion terms are equally important.
As before, we solve the steady problem in $x \in [0,1]$ by the implicit solver with $\kappa=0$, $1/2$, and $1/3$, 
for a series of grids with 15, 31, 63, 127 cells. 
The damping coefficient $\alpha$ is determined by Equation (\ref{alpha_kappa_fourth}) for a given $\kappa$ 
and the same $u_L$ and $u_R$ (i.e., the same $\kappa$) are used for both convective and diffusive fluxes.
Note that $u_L$ and $u_R$ are used to compute the damping term in the diffusive flux.
The analysis predicts that third-order accuracy is obtained only for $\kappa=1/3$. 
To verify the analysis, we computed the truncation errors as well, and the discretization errors for both
the cell-averaged and point-value exact solutions as described in the previous section.

  \begin{figure}[th!]
    \centering
      \hfill    
      \begin{center}
                \begin{subfigure}[t]{0.32\textwidth}
        \includegraphics[width=\textwidth]{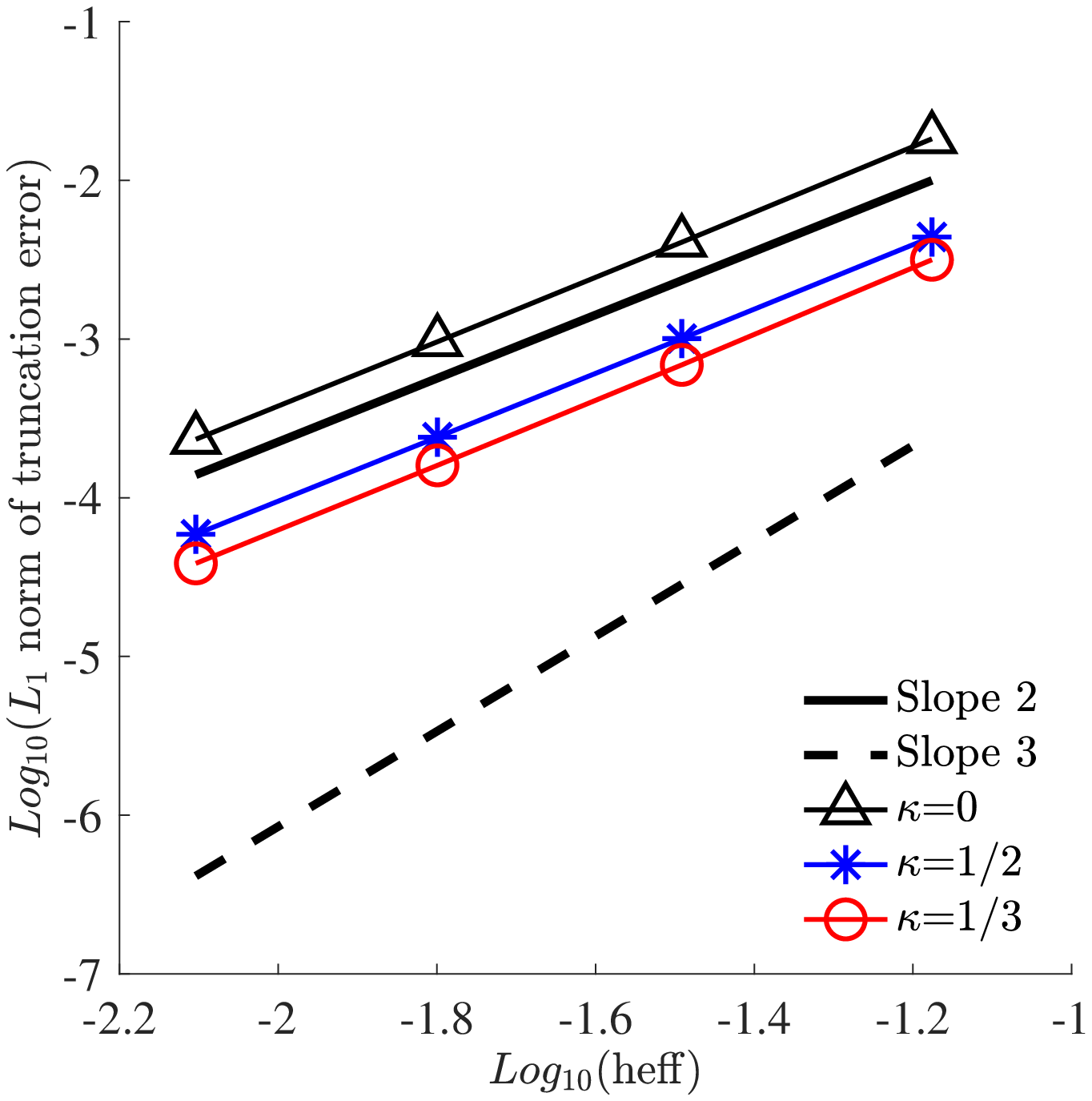}
          \caption{ $L_1( {\cal T}_p) $.}
       \label{fig:steady_adv_visc_te_p}
      \end{subfigure}
                \begin{subfigure}[t]{0.32\textwidth}
        \includegraphics[width=\textwidth]{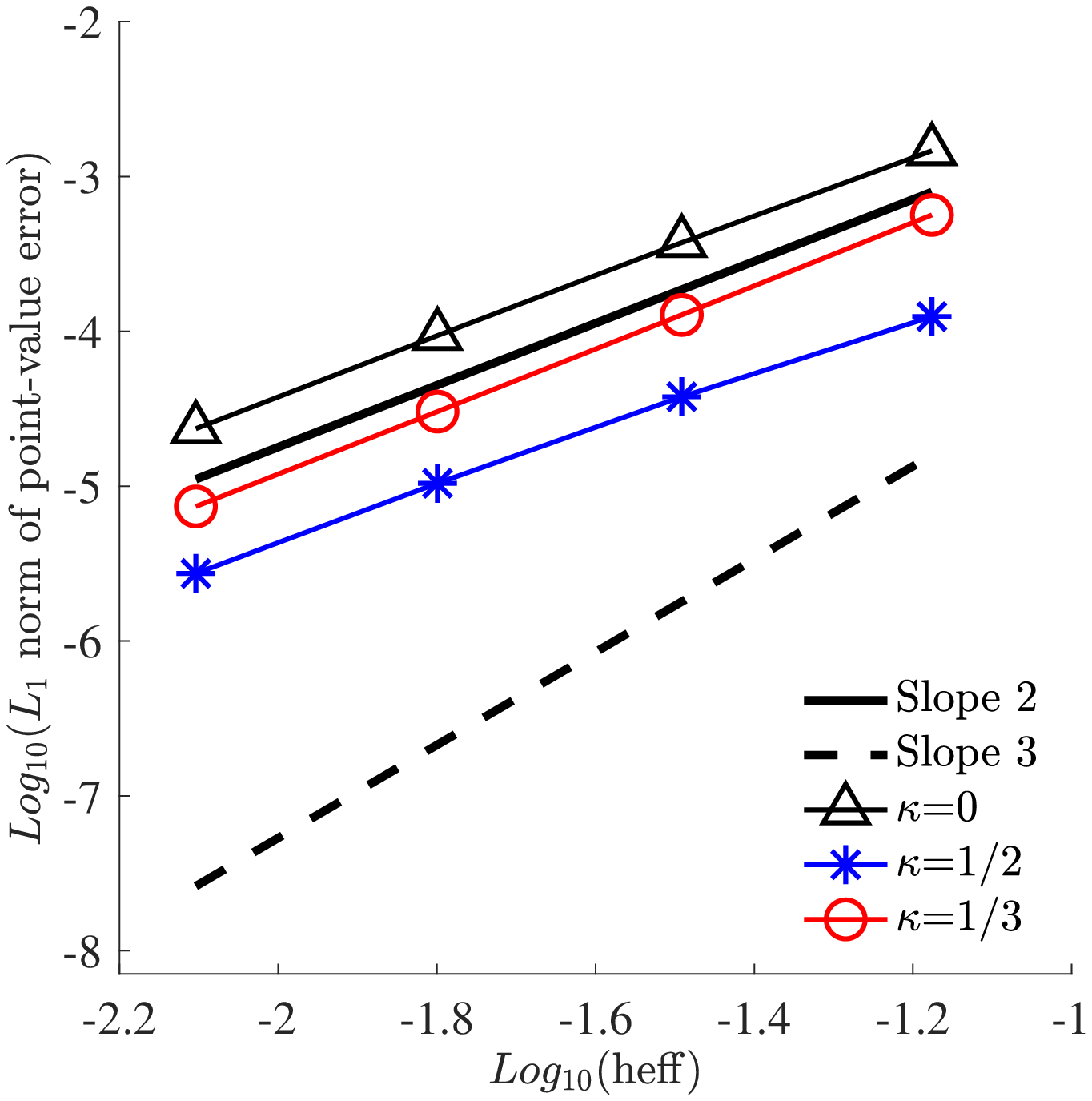}
          \caption{ $L_1( {\cal E}_p) $.}
       \label{fig:steady_adv_visc_err_p}
      \end{subfigure}
      \end{center}
      \hfill    
      \\
                \begin{subfigure}[t]{0.32\textwidth}
        \includegraphics[width=\textwidth]{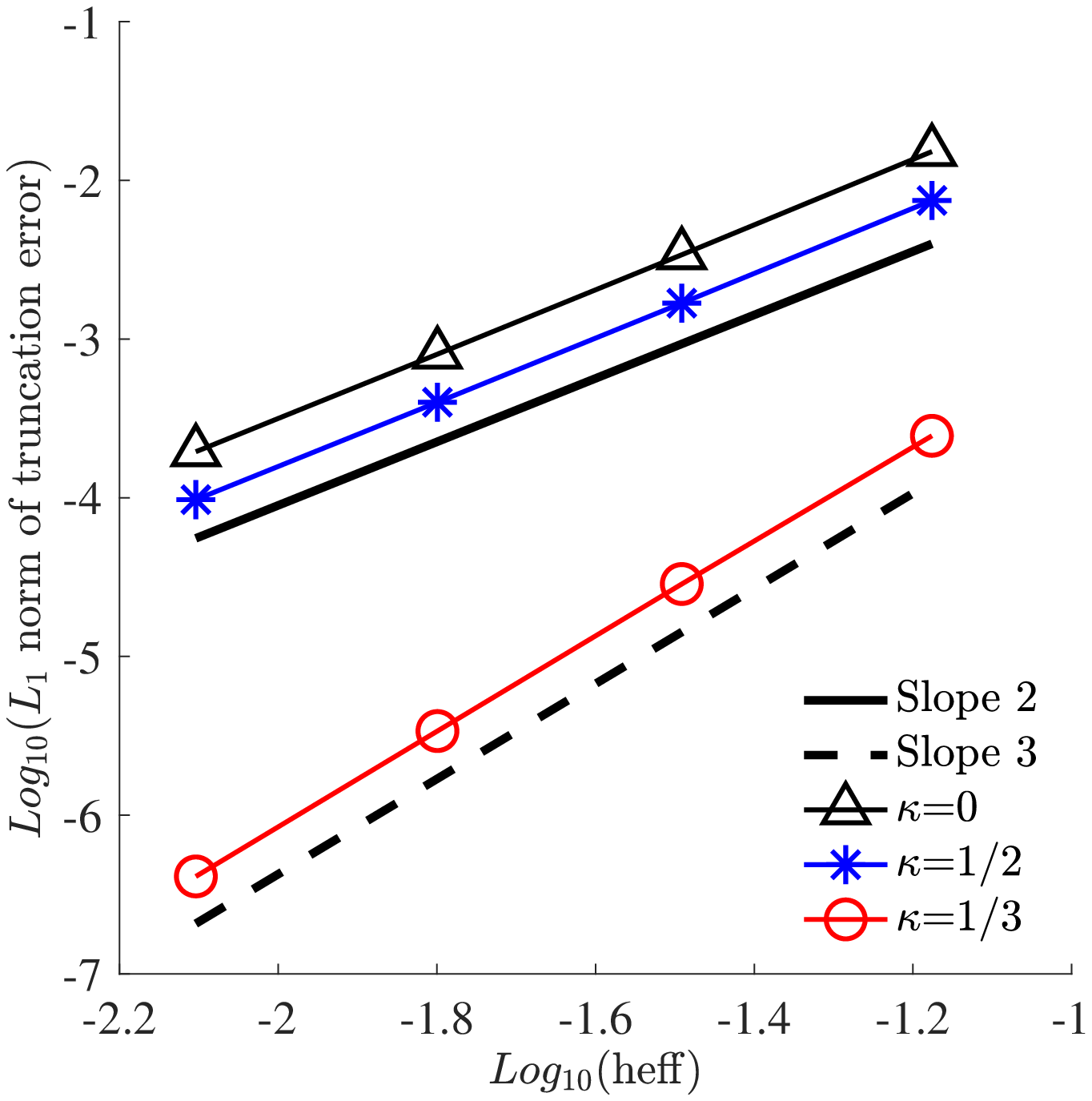}
          \caption{ $L_1( {\cal T}_c) $.}
       \label{fig:steady_adv_visc_te_c}
      \end{subfigure}
      \hfill
          \begin{subfigure}[t]{0.32\textwidth}
        \includegraphics[width=\textwidth]{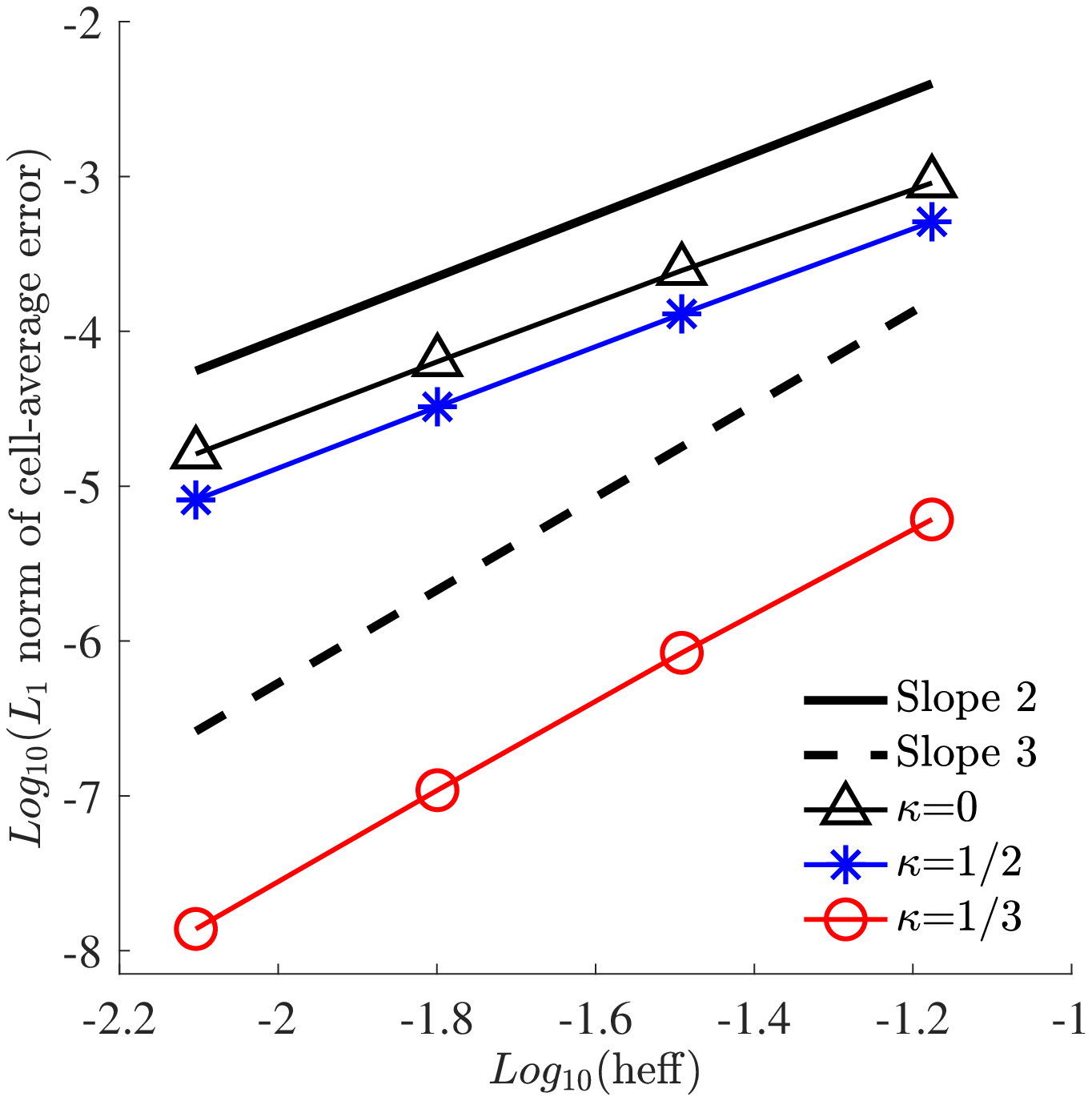}
          \caption{ $L_1( {\cal E}_c ) $.}
       \label{fig:steady_adv_visc_err_ca}
      \end{subfigure}
      \hfill
          \begin{subfigure}[t]{0.32\textwidth}
        \includegraphics[width=\textwidth]{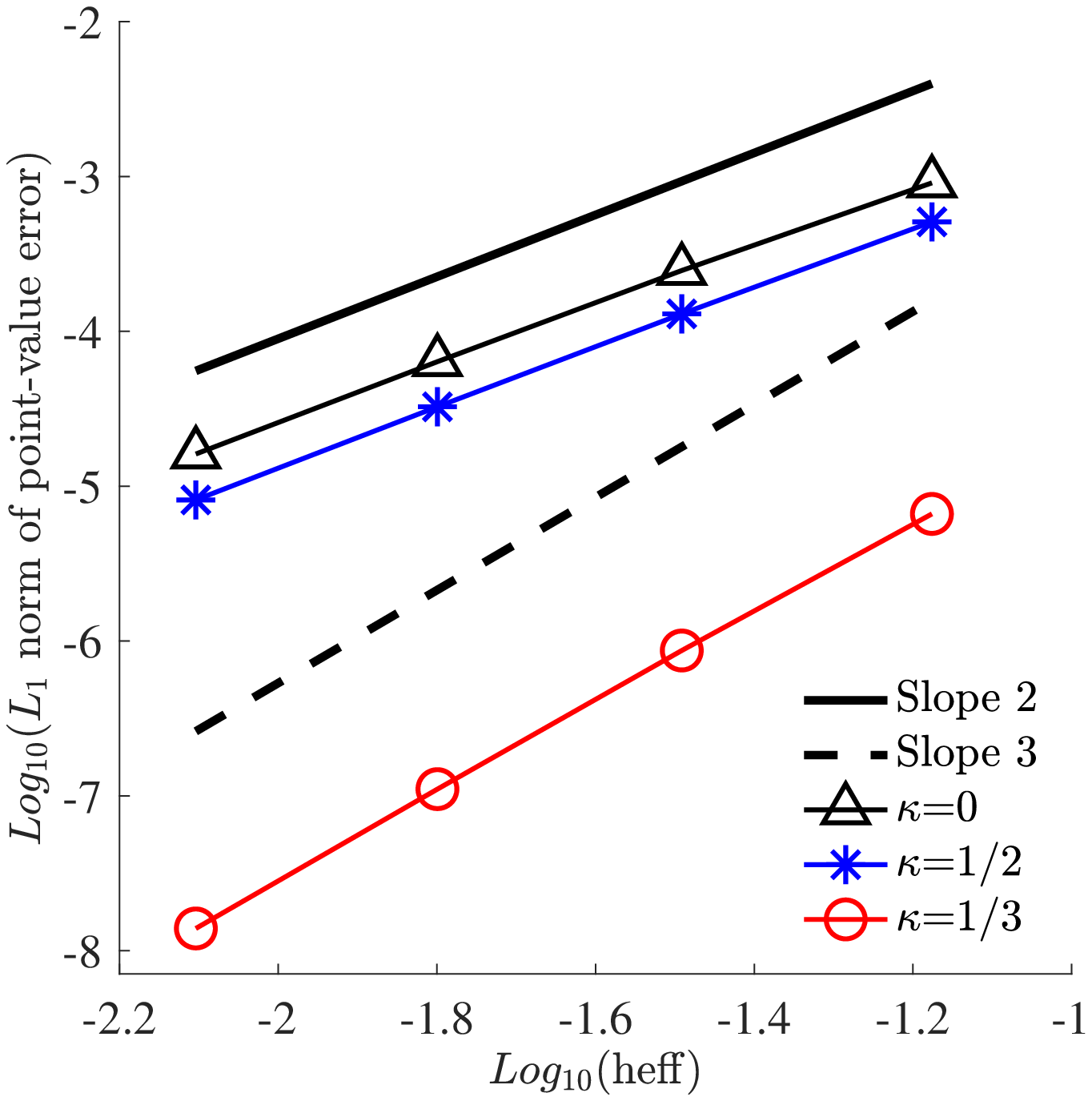}
          \caption{ $L_1( \hat{\cal E}_p ) $.}
       \label{fig:steady_adv_visc_err_pc}
      \end{subfigure}
      \hfill
            \caption{
\label{fig:steady_adv_visc_error}%
Truncation and discretization error convergence results for the case of the steady viscous Burgers equation.
The fourth-order finite-volume diffusion scheme with $ \alpha   =   \frac{  2  ( 4 - 9 \kappa ) }{ 3( 1- \kappa ) }$.
} 
\end{figure}
  \begin{figure}[th!]
    \centering
      \hfill 
                \begin{subfigure}[t]{0.32\textwidth}
        \includegraphics[width=\textwidth]{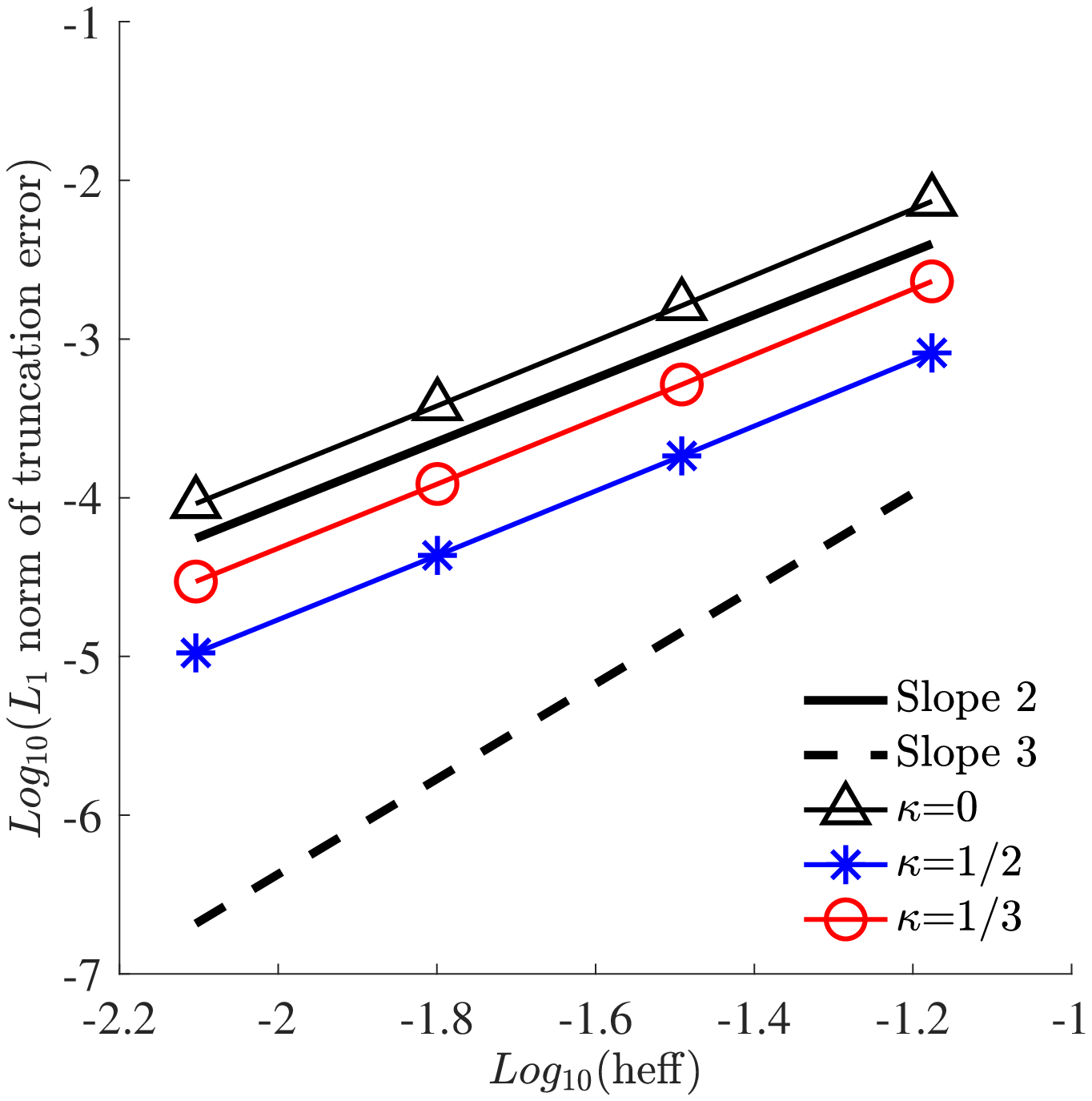}
          \caption{ $L_1( {\cal T}_c) $.}
       \label{fig:steady_adv_visc_te_c_alpha3o2}
      \end{subfigure}
      \hfill
          \begin{subfigure}[t]{0.32\textwidth}
        \includegraphics[width=\textwidth]{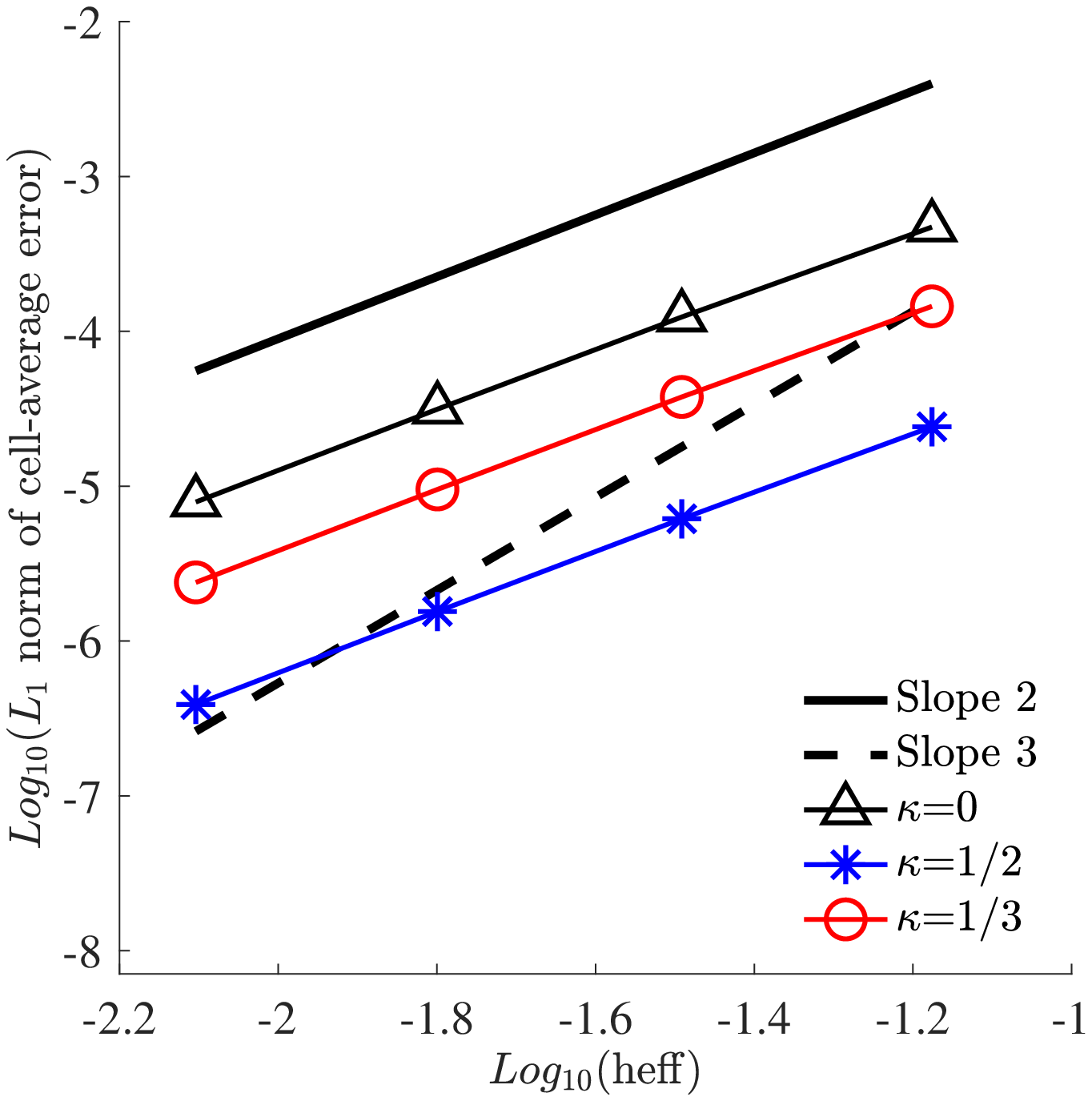}
          \caption{ $L_1( {\cal E}_c ) $.}
       \label{fig:steady_adv_visc_err_ca_alpha3o2}
      \end{subfigure}
      \hfill
          \begin{subfigure}[t]{0.32\textwidth}
        \includegraphics[width=\textwidth]{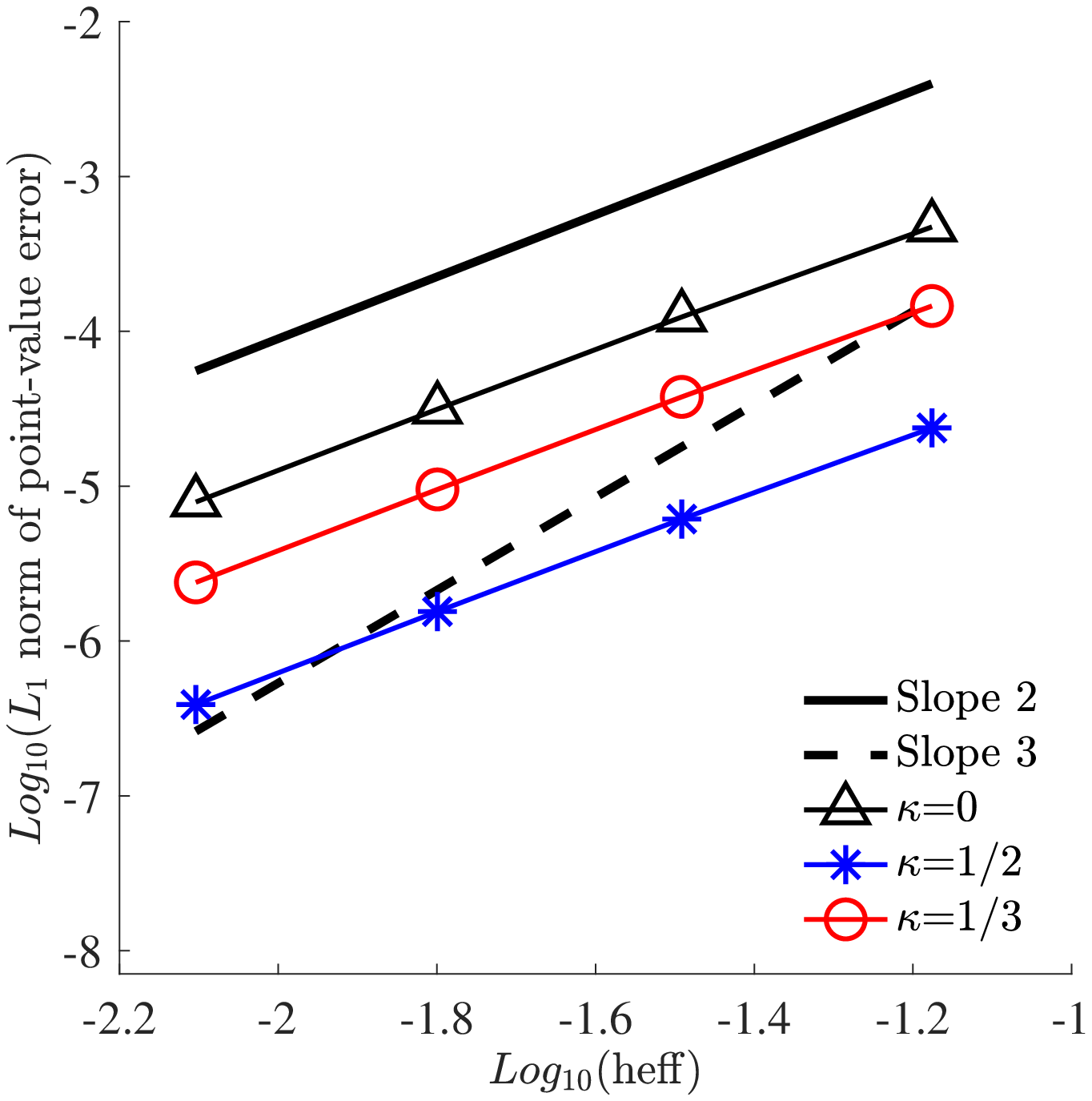}
          \caption{ $L_1( \hat{\cal E}_p ) $.}
       \label{fig:steady_adv_visc_err_pc_alpha3o2}
      \end{subfigure}
      \hfill
            \caption{
\label{fig:steady_adv_visc_error_alpha3o2}%
Truncation and discretization error convergence results for the case of the steady viscous Burgers equation.
The fourth-order finite-difference diffusion scheme (\ref{diffusion_scheme_wrong}) implemented 
as the alpha-damping scheme with $\alpha=3/2$.
} 
\end{figure}

Figure \ref{fig:steady_adv_visc_error} shows the results in terms of the point-value solution.
As shown in Figure \ref{fig:steady_adv_visc_te_p}, the truncation error $L_1( {\cal T}_p)$ 
is second-order for all choices of $\kappa$. Figure \ref{fig:steady_adv_visc_err_p} confirms that the discretization
errors are also second-order for all choices of $\kappa$. Therefore, unlike the previous case, third-order accuracy is not achieved with $\kappa=1/2$.
This is because the use of the point-value solution changes the diffusion scheme whereas it does not affect the cell-averaged forcing term.
As one can see, the condition (\ref{alpha_kappa_fourth}) for fourth-order accuracy has been derived for the diffusion scheme (\ref{fv__diffusion_01}) 
defined with the cell-averaged solutions. Once we replace the cell averages by the point values, the scheme involves the $O(h^2)$ error and loses
fourth-order accuracy. Then, the resulting convection-diffusion scheme is second-order accurate at best unless the diffusion term is negligibly small
compared with the convective term.
Therefore, the choice of the diffusion scheme is critically important to achieve third-order accuracy.

On the other hand, the scheme is third-order with the cell-averaged solution. 
Figure \ref{fig:steady_adv_visc_te_c} shows the truncation error convergence for $L_1( {\cal T}_c)$. 
As expected, the scheme achieves third-order accuracy with $\kappa=1/3$. Also, third-order accuracy 
is confirmed in the discretization errors. See Figures \ref{fig:steady_adv_visc_err_ca} and \ref{fig:steady_adv_visc_err_pc}.

Finally, we performed the computations with $\alpha = 3/2$, which corresponds to the incompatible diffusion scheme 
(\ref{diffusion_scheme_wrong}) when $\kappa=1/3$. Results are shown in Figure \ref{fig:steady_adv_visc_error_alpha3o2}. 
Clearly, as predicted in Section \ref{MUSCL_diffusion}, third-order accuracy is never achieved. Therefore, it is very important to choose the right diffusion scheme
in order to achieve third-order accuracy: the diffusion scheme must be a fourth-order finite-volume scheme with cell-averaged solution.

\section{Conclusions}
\label{conclusions}
 
We have provided a detailed truncation error analysis and proved that the MUSCL scheme is third-order accurate
with $\kappa=1/3$ for a nonlinear conservation law. Two key points have been emphasized: (1)the distinction of 
the cell average and the point value in the Taylor expansion of the solution and flux, and (2)the target operator
of the finite-volume discretization is the cell-averaged flux derivative, not the point value. 
It has been shown also that the $\kappa$-reconstruction scheme is exact for a cubic function on a uniform grid, when averaged across a face (i.e., averaging the reconstructed solutions
at the left and right of the face), and the same is true for the average of the fluxes evaluated with the reconstructed solutions. 
The importance of the diffusion scheme is also discussed: third-order accuracy is lost with an incompatible four-order diffusion scheme.
Third-order accuracy has been verified by numerical experiments for both unsteady and steady problems with Burgers' equation. 
For all the cases, it has been demonstrated that the MUSCL scheme gives third-order accurate cell-averaged solutions as well as third-order accurate 
point values at cell centers if the point values are accurately recovered from the third-order accurate cell averages. 
In a subsequent paper, we will discuss third-order accuracy of the QUICK scheme and clarify the reason that third-order accuracy was observed in the point value solution with $\kappa=1/2$ for a steady problem. 


\addcontentsline{toc}{section}{Acknowledgments}
\section*{Acknowledgments}

The author is grateful to Emeritus Professor Bram van Leer for illuminating discussions and helpful suggestions.
The author also would like to thank Jeffery A. White (NASA Langley Research Center) for valuable comments and discussions.
The author gratefully acknowledges support from Software CRADLE, part of Hexagon, the U.S. Army Research Office 
under the contract/grant number W911NF-19-1-0429 with Dr. Matthew Munson as the program manager, and 
 the Hypersonic Technology Project, through the Hypersonic Airbreathing Propulsion Branch of the NASA Langley
 Research Center, under Contract No. 80LARC17C0004.

\addcontentsline{toc}{section}{References}
\bibliography{../../bibtex_nishikawa_database}
\bibliographystyle{aiaa}

 
\end{document}